\RequirePackage{fix-cm}
\documentclass[smallextended]{svjour3}       % onecolumn (second format)
\smartqed  % flush right qed marks, e.g. at end of proof
\usepackage{graphicx}
\usepackage{multirow}
\usepackage{bibentry}
\usepackage{lineno}
\usepackage{setspace}
\begin{document}

\title{Testing a patient-specific \textit{in-silico} model to noninvasively estimate central blood pressure%\thanks{Grants or other notes
%about the article that should go on the front page should be
%placed here. General acknowledgments should be placed at the end of the article.}
}
%\subtitle{Do you have a subtitle?\\ If so, write it here}

\titlerunning{\textit{In-silico} central blood pressure estimation}        % if too long for running head

\author{Caterina Gallo \and Joakim Olbers \and Luca Ridolfi \and Stefania Scarsoglio \and Nils Witt}

%\authorrunning{Short form of author list} % if too long for running head

\institute{C. Gallo \and S. Scarsoglio \at
              Department of Mechanical and Aerospace Engineering,
              Politecnico di Torino, Torino, Italy \\
              \email{caterina.gallo@polito.it} \and J. Olbers and N. Witt \at
              Department of Clinical Science and Education, Karolinska Institutet, Division of Cardiology, Södersjukhuset, Stockholm, Sweden \and L. Ridolfi \at Department of Environmental, Land and Infrastructure Engineering, Politecnico di Torino, Torino, Italy
}

\date{Received: date / Accepted: date}
% The correct dates will be entered by the editor

\maketitle
%\linenumbers
%\doublespacing

\section*{Abstract}

\textbf{Purpose}: To show some preliminary results about the possibility to exploit a cardiovascular mathematical model - made patient-specific by noninvasive data routinely measured during ordinary clinical examinations - in order to obtain sufficiently accurate central blood pressure (BP) estimates.

\textbf{Methods}: A closed-loop multiscale (0D and 1D) model of the cardiovascular system is made patient-specific by using as model inputs the individual mean heart rate and left-ventricular contraction time, weight, height, age, sex and mean/pulse brachial BPs. The resulting framework is used to determine central systolic, diastolic, mean and pulse pressures, which are compared with the beat-averaged invasive pressures of 12 patients aged 72$\pm$6.61 years.

\textbf{Results}: Errors in central systolic, diastolic, mean and pulse pressures by the model are 4.26$\pm$2.81 mmHg, 5.86$\pm$4.38 mmHg, 4.98$\pm$3.95 mmHg and 3.51$\pm$2.38 mmHg, respectively.

\textbf{Conclusion}: The proposed modeling approach shows a good patient-specific response and appears to be potentially useful in clinical practice. However, this approach needs to be evaluated in a larger cohort of patients and could possibly be improved through more accurate oscillometric BP measurement methods.

% \PACS{PACS code1 \and PACS code2 \and more}
% \subclass{MSC code1 \and MSC code2 \and more}

%\end{abstract}

\section{Introduction}
\label{intro}

Automatic brachial BP monitoring is routinely used in clinical practice to get an easily obtainable and noninvasive measurement of arterial BP. However, it is widely accepted that brachial BP is not a good estimation of central BP, mainly because of the amplification in the pressure waveforms from the ascending aorta to the stiffer peripheral arteries \cite{McEniery1}. Considering that target organs are more subjected to central than brachial BP \cite{Kollias}, simple estimation of central BP is expected to be helpful in diagnostics and clinical decision making. In support of this, it has been demonstrated that the addition of central BP measurement to conventional brachial BP measurement may identify individuals with an elevated central BP but a brachial BP in the normal or high normal range \cite{McEniery2}. These individuals may have an increased risk for cardiovascular events not reflected by their brachial BP. Moreover, the fact that antihypertensive medications can produce different effects on central BP but comparable effects on brachial BP \cite{Williams} makes the idea to clinically use central BP attractive.

Cardiac catheterization still represents the most accurate method to evaluate central BP. However, it cannot be appropriately employed in routine clinical setting since it is an invasive procedure, technically demanding and time consuming. Instead, a number of noninvasive devices have been proposed and tested to provide central BP estimates from distal pressure signals and through a variety of calibration techniques \cite{Yao}. So far, the incremental value of noninvasive central BP compared to brachial BP in the prediction of serious cardiovascular events has not been unequivocally demonstrated, with different studies coming up to opposite conclusions \cite{Sharman}. Thus, it seems that the accuracy of current noninvasive methods for estimation of central BP is not always sufficient to confirm superiority over brachial BP \cite{Laurent}. In this context, the need emerges to identify the most accurate solutions to noninvasively estimate central BP, before further investigating if brachial or central BP is a better prognostic parameter to be adopted in the future.

We present some preliminary results about the possibility to estimate the individual central BP by a patient-specific multiscale mathematical model of the cardiovascular system. In the last few decades, several models of the cardiovascular system have been proposed and tested to shed light into the pathogenesis of cardiac and vascular diseases \cite{Taylor}, sustain the design of medical devices (e.g., stents and valve prostheses) \cite{Morris}, support teaching and training activities \cite{Abram}, and prognosticate the effects of potential therapeutic plans \cite{Smith}. From a clinical perspective, the modeling approach is demonstrating an attracting tool. In fact, it is less expensive than \textit{in-vivo} studies and leads to reliable results. Moreover, it can be used to isolate the role played by specific pathologies and provide more information about the whole hemodynamic picture.

Cardiovascular models are beginning to be tailored on each patient by the so-called patient-specific approaches, which depend on data (e.g., vascular geometry, heart data, cardiac electrical activity, etc.) measured on the examined patient. The latter modeling approach is defined \textit{patient-specific} and is typically based on the adoption of the exact patient-specific vascular geometry derived from scans. In the following, we will use the term \textit{patient-specific} in a more general meaning, i.e., to underline the effort to adapt, even partially, the model to a specific patient, as already presented by other authors \cite{Passera,Guala1}. Indeed, to make the model patient-specific, we did not use patient-specific vascular data, but we adopted empirical rules (extracted from measurements on large cohort of individuals) introducing the patient-specific personal and anthropometric data.

This study aims at showing the potential use of the \textit{in-silico} approach to noninvasively evaluate central BP on specific patients. Despite the preliminary nature of our results, they represent an invitation to continue working on patient-specific models yielding increasingly accurate central BP estimations, apart from an extended set of patient-specific hemodynamic parameters.

\section{Materials and Methods}
\label{sec:1}

Noninvasive patient-specific data, consisting of anthropometric and clinical measures of 12 subjects, were used as model inputs to compute the individual aortic BPs. Simulated pressures were then compared with the beat-averaged values of the invasive ascending aorta pressure signals, quantifying the errors produced by the proposed modeling approach in terms of central systolic, diastolic, pulse and mean pressures.

\subsection{Patient-specific data}
\label{subsec:1}

Anthropometric and clinical measures adopted as model inputs include sex ($S$), age ($AGE$), weight ($W$), height ($H$), mean heart rate ($HR$), mean left ventricular contraction time ($T_{vc}$), mean and pulse brachial BPs ($P_{m_b}$ and $PP_b$). $HR$ and $T_{vc}$ were extracted from the ECG, while $PP_b$ and $P_{m_b}$ were calculated from the systolic ($Ps_b$) and diastolic ($Pd_b$) brachial BPs ($PP_b=Ps_b-Pd_b$ and $Pm_b=Pd_b+PP_b/3$). The latter were noninvasively measured from the left upper arm through automatic oscillometric recording, simultaneously to the invasive pressure at the level of the ascending aorta. Measurements were performed using a Philips IntelliVue MMS X2 bedside monitoring system with a Philips Easy Care Adult cuff (Philips Healthcare, Andover, Ma). Appropriate cuff size was used according to the individual circumference of the left upper arm.

Invasive pressure signals derived from a prospective clinical study examining intra-arterial BP in patients undergoing routine coronary angiography \cite{Olbers} (the trial was approved by the regional ethics committee). Intra-arterial pressure recordings were acquired from 12 patients with sinus rhythm, using a fluid-filled catheter system. Five French 100 cm long right coronary diagnostic catheters, connected to a pressure transducer-equipped manifold with two taps (NAMIC, Navilyst Medical Inc, Marlborough, MA, USA), were used. Pressure signals were recorded using the RadiAnalyzer Xpress unit (St Jude Medical, St Paul, MN, USA) for digital storage. No data on the wave form sampling rate has been provided by the manufacturer. The catheter system was flushed with saline before starting the measurement for elimination of air bubbles from tubes and connecting parts. After zeroing to air, the pressure transducer was adjusted to estimated left atrial level. No specific test for frequency response of the catheter system was carried out, but the natural frequency of the pressure transducer itself has been specified by the manufacturer to be 200 Hz. Right radial, right brachial and ascending aorta pressure signals were sequentially recorded by advancing a diagnostic catheter from the right radial to the right brachial to the central site (ascending aorta). Appropriate location of the catheter tip was confirmed by fluoroscopy. At least 15 cardiac cycles were saved at each location for all subjects. Intra-arterial BP recording was initiated at the start of cuff deflation in the left arm, thus making invasive and non-invasive BP recordings simultaneous.

\subsection{Patient-specific mathematical model}
\label{subsec:2}

The modeling approach implemented in this study is a closed-loop multiscale
mathematical model of the cardiovascular system. The latter originates from a physically-based 1D representation of the systemic arterial tree \cite{Guala1}, which was previously used to study the aging process \cite{Guala2} and the impact of atrial fibrillation \cite{Scarsoglio1,Scarsoglio2}. A 1D model is adopted to reproduce the arterial tree hemodynamics, rather than a 0D model, because it allows us to properly describe the reflection and propagation phenomena of pressure and flow waves. Thus, much more hemodynamic information are available at a reasonable computational cost. Moreover, in lumped models, vascular lengths and diameters are not explicitly modeled, making extremely difficult to apply patient-specific adjustments on vascular geometry.

The 1D model of the arterial tree is integrated with a 0D description of the remaining portions of the cardiovascular system, that is the systemic microcirculation and venous return, the heart and pulmonary circulation, and the short-term baroreflex mechanism to maintain homeostasis. The resulting \textit{in-silico} model, which adequately describes the physiological hemodynamic behavior of a reference healthy subject, was validated in heart pacing and open-loop response \cite{Gallo1}, and exploited to inquire into the effects of long duration spaceflights on the cardiovascular system \cite{Gallo2}. A schematic representation of the model and a summary of the equations used to reproduce its constitutive parts are given in the Online Resource. Further details on the 1D-0D model can be found elsewhere \cite{Guala1,Guala2,Scarsoglio1,Scarsoglio2,Gallo1,Gallo2}.

The reference multiscale model corresponds to a generic healthy subject with the following characteristics: $HR_{ref}=75$ bpm, $T_{vc_{ref}}=0.27$ s, $W_{ref}=75$ kg, $H_{ref}=175$ cm, $AGE_{ref}=25$ years, $S_{ref}$=man, $P_{m_{b_{ref}}}=88$ mmHg and $PP_{b_{ref}}=67$ mmHg. Notice that $HR$, $T_{vc}$ and $P_{m_b}$ appear explicitly in the model, while $AGE$, $W$, $H$, $S$ and $PP_b$ are implicit in the geometrical and cardiac/vascular mechanical properties.

\begin{figure*}[!h]
\begin{center}
\includegraphics*[width=0.95\textwidth]{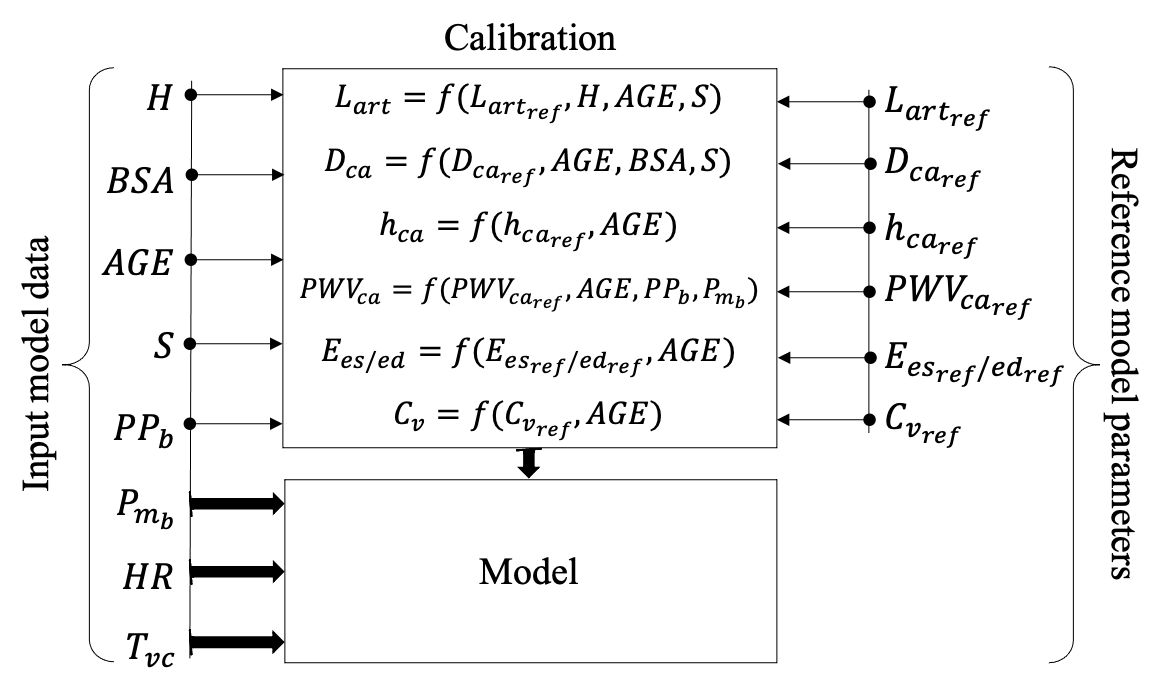}
\caption{}
\label{Figure1}
\end{center}
\end{figure*}

To make the model patient-specific, we used $HR$, $T_{vc}$, $W$, $H$, $AGE$, $S$, $P_{m_b}$ and $PP_b$ as model input data depending on the patient characteristics. In fact, it is widely accepted in literature that anthropometric ($W$ and $H$) and personal data ($AGE$ and $S$) are crucial hemodynamic determinants, together with time-averaged ECG parameters and pressure level, which can be different among people with the same anthropometric and personal data. Coherently to the explicit or implicit occurrence of the different model input data in the reference model, patient-specific values of $HR$,  $T_{vc}$ and $P_{m_b}$ were directly introduced in the model, while individual values of $W$, $H$, $AGE$, $S$ and $PP_b$ were used in empirical relationships to adapt the arterial geometry and cardiac/vascular mechanical properties to the specific patient characteristics (see Fig. \ref{Figure1}). In this way, we were able to match the reference model to each patient considered. In particular, arterial lengths, diameters and thicknesses, as well as cardiac, arterial and venous compliances of the reference subject - all marked with the subscript \textit{ref} and provided in a separate publication \cite{Gallo2} - were adapted to the specific patient by suitable empirical relationships, as described below. In order to compare our modelling results with pre-existing clinical data on old and not perfectly healthy patients (already measured at the Department of Clinical Science and Education, Karolinska Institutet, Division of Cardiology, Södersjukhuset, Stockholm, Sweden), we considered empirical relationships derived from large cohorts and representative of white people affected by cardiovascular or cardiovascular-related problems, such as hypertension, diabetes and atrial fibrillation. To immediately recognize the direct dependence of each vascular property to the corresponding input parameters, the following empirical relationships will be expressed in dimensional form, as they are typically found in literature. An effort to restate them in dimensionless form will be done in future (more comprehensive) works.

\paragraph{Arterial lengths.}
Arterial path length depends on $H$, which then influences the resultant hemodynamic behavior \cite{Langenberg1,Langenberg2,London,Smulyan}. In shorter people, in fact, the earlier arrival of the reflected waves to the heart during systole causes an increase in both systolic and pulse pressures, which are responsible for a rise in the left-ventricular work and stress at the same mean pressure. To take into account these phenomena, we adjusted arterial lengths with $H$, scaling them according to the patient-specific $H$. Patient-specific arterial lengths, $L_{art}$, were modified from reference lengths, $L_{art_{ref}}$ \cite{Guala1}, as

\begin{equation}
L_{art}=L_{art_{ref}}\frac{H}{H_{ref}}+c_1(AGE-AGE_{ref}),
\end{equation}

\noindent with the term $c_1(AGE-AGE_{ref})$ introducing the effect of $AGE$ on arterial lengths. In fact, some arterial tracts also elongate with $AGE$, with different relationships proposed in literature \cite{Rylski,Sugawara}. We modified the reference aortic lengths with $AGE$ according to the regression coefficients, $c_1$, by Rylski et al. \cite{Rylski}, reported in Table \ref{Tab1}, while maintained the other arterial lengths constant with $AGE$. Notice that Rylski’s coefficients (which are sex-dependent) are provided for a change in mm per year and per m$^2$ of body surface area, $BSA$. Thus, coefficients $c_1$ have to be multiplied for the patient-specific $BSA$, which we calculated from $H$ and $W$ with the Du Bois’s formula \cite{Du_Bois}: $BSA=0.20247H^{0.725}W^{0.425}$ ($H$ and $W$ are given in m and kg, respectively).

\paragraph{Arterial diameters.}
It is well known that diameters of elastic arteries widen with $AGE$, while muscular arteries are not subject to enlargement during normal aging \cite{Man}. However, if there is a general agreement about the age-induced rise in the aortic and common carotid arterial diameters, there is some controversy regarding the changes with $AGE$ in the diameters of some muscular arteries, such as the brachial, radial and common femoral arteries. In fact, some authors report decreases in the diameters of these arteries \cite{Boutouyrie}, while others sustain the opposite trend \cite{Green}. This controversy could be due to the existence of a transition zone between the elastic and muscular arterial behavior at more distal arterial sites, as observed by \cite{Bjarnegard}, who monitored the age variations in the brachial artery diameter both proximally and distally. Considering the contrasting behavior observed at more peripheral sites, we  modeled the age effects on aortic and carotid diameters only. In particular, the role of $AGE$ was reproduced through the regression coefficients $c_2$, given in Table \ref{Tab2}, by Rylski et al. \cite{Rylski} along the aorta and by Kamenskiy et al. \cite{Kamenskiy} for the common carotid arteries, respectively. All these coefficients are sex-dependent, and refer to a change in mm per year and per m$^2$ of $BSA$ by Rylsky et al. \cite{Rylski}, and for a change in mm per year by Kamenskiy et al. \cite{Kamenskiy}. Since an explicit dependence of the aortic diameters with $BSA$ was also considered (as described later in the text), age-dependent coefficients by Rylsky et al. \cite{Rylski} were multiplied for the reference $BSA$, $BSA_{ref}$=1.90 m$^2$, calculated through the Du Bois's formula.

Arterial diameters also change with both $W$ and $H$, whose combined effect can be taken into account by either the body mass index ($BMI=W/H$) or $BSA$. In this study we used $BSA$ as body size variable, considering that it was proved to be better correlated to some aortic diameters than $BMI$ \cite{Wolak}. We here introduced the effects of $BSA$ at aortic and carotid level only, where the role of $BSA$ has been largely assessed \cite{Rylski,Kamenskiy,Wolak,Davis,Fleischmann,Krejza,Ruan}. The effect of $BSA$ on the aortic and common carotid arteries diameters were quantified through the regression coefficients $c_3$ by Davis et al. \cite{Davis} and Krejza et al. \cite{Krejza}, respectively, both indicated in Table \ref{Tab2} and expressing different changes for women and men.

Thus, patient-specific carotid and aortic diameters, $D_{ca}$, were determined from the corresponding reference values, $D_{ca_{ref}}$ \cite{Guala1}, as

\begin{equation}
D_{ca}=D_{ca_{ref}}+c_2(AGE-AGE_{ref})+c_3(BSA-BSA_{ref}).
\end{equation}

\paragraph{Arterial thicknesses.}
Aortic and common carotid arteries thicken with $AGE$ \cite{Man,Howard,Rashid}, while there are contrasting results at other arterial locations. Thus, only patient-specific aortic and common carotid thicknesses, $h_{ca}$, were modified with $AGE$ from reference values, $h_{ca_{ref}}$ \cite{Guala2}. Namely,

\begin{equation}
h_{ca}=h_{ca_{ref}}+c_4(AGE-AGE_{ref}),
\end{equation}

\noindent where $c_4$ was taken by Virmani et al. \cite{Guala2,Virmani} along the aorta and according to Howard et al. \cite{Howard} and Rashid et al. \cite{Rashid} for the common carotid arteries. In particular, we set coefficients $c_4$ equal to 0.0040 mm/y for the ascending aorta, 0.0092 mm/y for the descending thoracic aorta, 0.0085 mm/y for the suprarenal abdominal aorta, 0.0144 mm/y for the subrenal abdominal aorta, and 0.010 mm/y for the common carotid arteries, neglecting any sex difference and side-to-side effects for left and right carotids.

\paragraph{Arterial compliance.}
Elastic arteries stiffen with $AGE$, leading to a decrease in arterial compliance. It was demonstrated that arterial stiffening is mainly due to the fatigue and successive rupture of the median elastic lamellae, which are expected to fracture after about 8$\cdot$10$^8$ cycles (e.g., 30 years with a mean HR of 70 bpm) \cite{Orourke1,Orourke2}. Carotid-femoral pulse wave velocity ($PWV$) is a surrogate index of aortic stiffness \cite{VanBortel}, with several studies on its changes with aging available in literature. However, carotid-femoral $PWV$ does not take into account the influence of proximal aorta, with only regional $PWV$s adequately mapping the differential stiffness along the aorta and at other arterial locations. Numerous authors have measured the variations in $PWV$s with $AGE$ along the aorta \cite{Hickson,Redheuil,Rogers} and common carotids \cite{Borlotti,Vritz}, finding quadratic and linear relationships, respectively. Alterations in $PWV$s with $AGE$ at other arterial sites were found negligible or absent and were here neglected. We adapted the patient-specific carotid and aortic compliances through the relative patient-specific pulse wave velocities ($PWV_{ca}$s). In fact, mechanical properties of 1D arteries are specified through coefficients $B_{1-5}$ of the constitutive equation for pressure (see Eq. T3 in the Online Resource), which are function of the local $PWV$s. Aortic $PWV$s ($PWV_a$s) were calculated from reference values, $PWV_{a_{ref}}$ \cite{Guala1}, as

\begin{equation}
PWV_{a}=PWV_{a_{ref}}+[c_5(AGE-AGE_{ref} )+c_6(AGE^2-AGE_{ref}^2)]a+b,
\label{PWVa}
\end{equation}

\noindent where $c_5$ and $c_6$, provided in Table \ref{Tab1}, are those by Hickson et al. \cite{Hickson}. Carotid $PWV$s ($PWV_{c}$s), instead, were evaluated from reference values, $PWV_{c_{ref}}$ \cite{Guala1}, as

\begin{equation}
PWV_{c}=PWV_{c_{ref}}+c_7(AGE-AGE_{ref})a+b,
\label{PWVb}
\end{equation}

\noindent with $c_7=0.0538$ m/s/year given by Vriz et al. \cite{Vritz}. In equations (\ref{PWVa}-\ref{PWVb}), coefficient $a$ and term $b$ allow one to include the role of the pressure level ($PP_{b}$ and $P_{m_{b}}$) in the patient-specific arterial compliances \cite{Kim,Mattace_Raso}. Notice that reference values of local $PWV$s are calculated from the quasi-linear form of the system of equations solving the 1D arteries (Eqs. T1, T2 and T3 in the Online Resource):

\begin{equation}
PWV_{a/c_{ref}}=\sqrt{\frac{A}{\rho}\left(B_2+2B_3A+3B_4A^2\right)+\left(\beta-1\right)\beta\frac{Q^2}{A^2}},
\label{PWVc}
\end{equation}

\noindent where variables are defined in Table S2 of the Online Resource.

\paragraph{Cardiac compliances.}
Since ventricular stiffening increases with $AGE$ \cite{Redfield}, we corrected patient-specific end-systolic, $E_{es}$, and end-diastolic, $E_{ed}$, left-ventricular elastance values, starting from the reference values \cite{Guala1} ($E_{es_{ref}}$ and $E_{ed_{ref}}$, respectively), as

\begin{equation}
E_{es}=E_{es_{ref}}+c_{8}(AGE-AGE_{ref})
\end{equation}

\noindent and

\begin{equation}
E_{ed}=E_{ed_{ref}}+c_{9}(AGE-AGE_{ref}).
\end{equation}

\noindent Coefficient $c_{8}$ was set to have an increase of 1\% and 0.5\% per year for women and men, respectively, according to Redfield et al. \cite{Redfield}. $c_{9}$ was instead reasonably taken equal to $c_{8}E_{ed_{ref}}/E_{es_{ref}}$.

\paragraph{Venous compliances.}
Venous compliances, as arterial ones, reduce with $AGE$ \cite{Greaney,Olsen}. We evaluated the patient-specific venous compliances, $C_{v}$, from the reference values, $C_{v_{ref}}$ \cite{Gallo2}, through the

\begin{equation}
C_{v}=C_{v_{ref}}+c_{10}(AGE-AGE_{ref}),
\end{equation}

\noindent with $c_{10}$ chosen in order to have a linear reduction in $C_{v_{ref}}$ of 50\% from 25 to 80 years \cite{Olsen}.

\subsection{Sensitivity analysis} \label{sensitivity_analysis}

In order to quantify the role of the different patient-specific input data to the individual central pressure values, we performed a sensitivity analysis. Considering an input parameter $X$ and an output parameter $Y$, the sensitivity of $Y$ to $X$ is defined as

\begin{equation}
s=\left(\frac{Y'-Y}{Y}\right)\left(\frac{X}{X'-X}\right),
\end{equation}

\noindent where $Y'$ is the modified output parameter obtained with the modified input parameter $X'$ \cite{Mynard_th}. Based on this definition, negative values of $s$ imply that an increase in $X$ causes a decrease in $Y$ and a decrease in $X$ causes an increase in $Y$. Moreover, $|s|$ values higher (smaller) than 1 indicate that the input variability introduced through $X$ is amplified (damped) in $Y$. Here, we imposed an increase of 25\% to all the patient-specific input data, that is $X'=X+0.25X$.

\begin{figure*}[!h]
\begin{center}
\includegraphics*[scale=0.4]{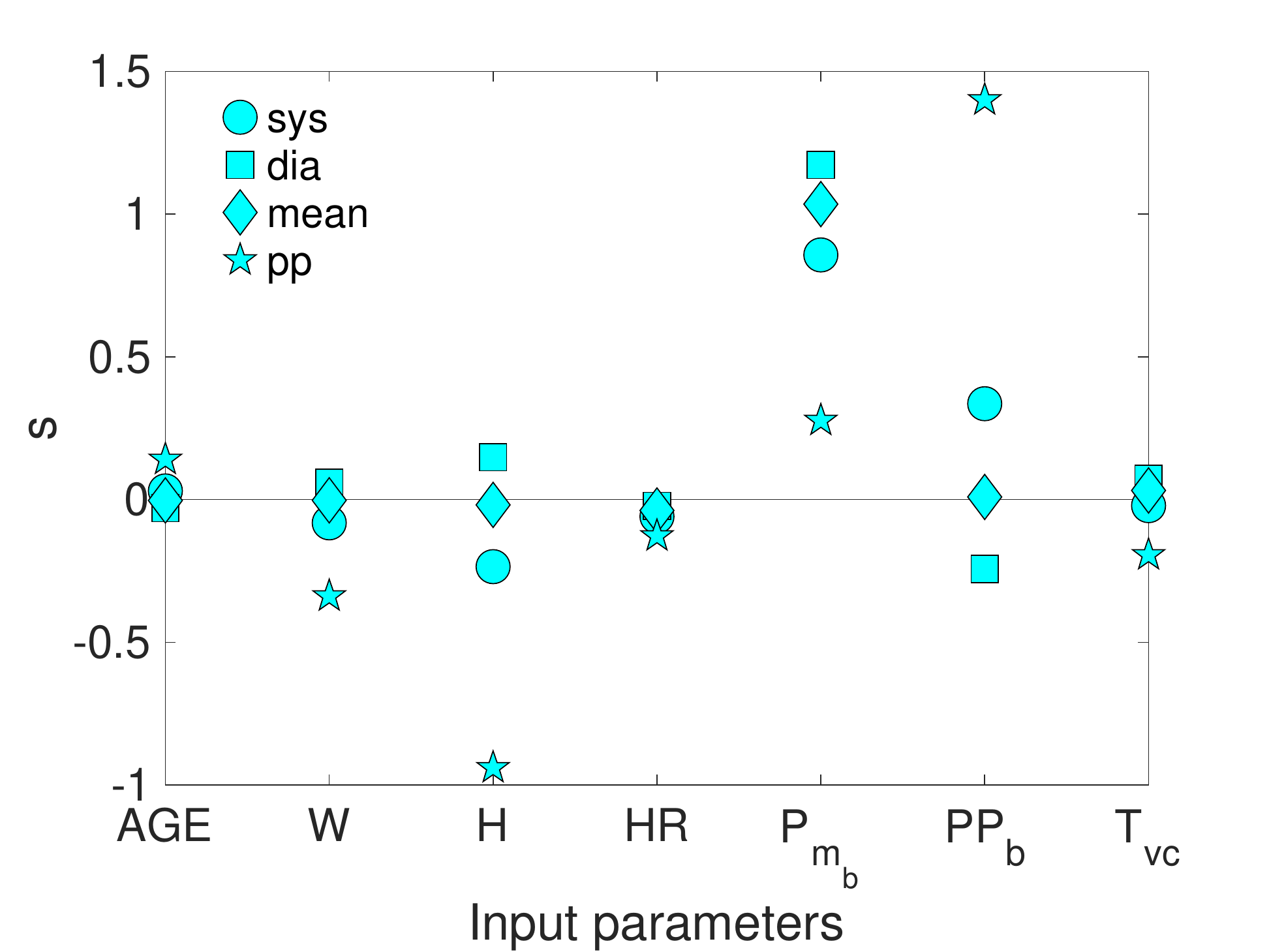}
\caption{}
\label{Figure2}
\end{center}
\end{figure*}

Fig. \ref{Figure2} shows the sensitivities of central systolic (sys), diastolic (dia), mean (mean) and pulse (pp) pressures to $AGE$, $W$, $H$, $HR$, $T_{vc}$, $P_{m_b}$ and $PP_b$ for men. Results for women are very similar to the ones for men (they are reported in Table S4 in the Online Resource). From Fig. \ref{Figure2}, it emerges that $P_{m_b}$ is the only input parameter having a not negligible impact on all the output parameters, and $H$ and $PP_b$ are the input parameters with the greatest impact on central pp. The sensitivity of central pp to $H$ is negative (-0.94), while the sensitivity of central pp to $PP_b$ is positive (1.40). The other input parameters ($AGE$, $W$, $HR$ and $T_{vc}$) result to be less effective on the output parameters, with the smallest sensitivities for central mean and the highest for central pp. From Fig. \ref{Figure2} it also appears that the variability introduced through $P_{m_b}$ is amplified in central dia and mean (and damped in central sys and pp), and the variability introduced through $PP_b$ is amplified in central pp (and damped in central sys, dia and mean). The variability associated to all the other input parameters results to be damped. Thus, according to Fig. \ref{Figure2}, specific brachial BPs and $H$ measurements are the most influential input data on the model outputs, although the majority of the chosen patient-specific input data are proven to be important to match central pp.

\section{Results}

Anthropometric and clinical data, presence of comorbidities, like diabetes and ischemic heart disease (IHD), and smoking status of patients are reported in Table \ref{Tab3}.

Mean, $\mu$, standard deviation, $\sigma$, and coefficient of variation ($cv=\sigma/\mu$) values of systolic and diastolic pressures recorded at the three measurement sites for all the patients are indicated in Table \ref{Tab4}. Based on systolic BP transmission from central-to-peripheral sites, Picone et al. \cite{Picone} individuated four BP phenotypes: (phenotype I) both central-to-brachial and brachial-to-radial systolic BP increase ($\geq$5 mmHg), (phenotype II) only aortic-to-brachial systolic BP increase, (phenotype III) only brachial-to-radial systolic BP increase, (phenotype IV) neither aortic-to-brachial nor brachial-to-radial systolic BP increase. Considering the invasive measurements of systolic BP at the ascending aorta, brachial and radial arteries on the 12 patients (all indicated in Table \ref{Tab4}), we recognized all the four BP phenotypes identified by Picone et al. \cite{Picone}. The phenotype associated to each patient is reported in Table \ref{Tab4}.

\begin{figure*}[!h]
\begin{center}
\includegraphics*[width=0.9\textwidth]{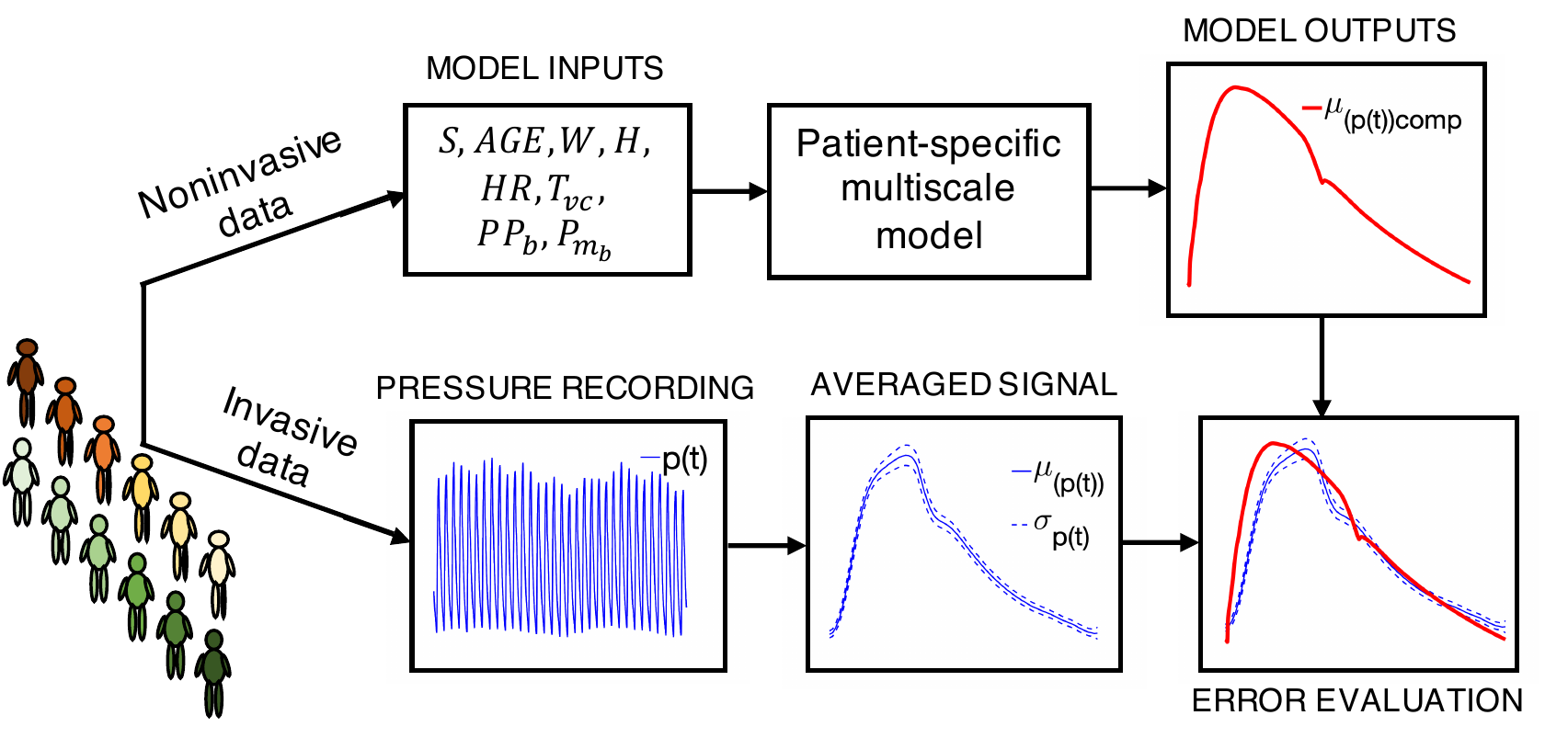}
\caption{}
\label{Figure3}
\end{center}
\end{figure*}

To test the reliability of the patient-specific multiscale model to estimate the individual central BP, the procedure schematized in Fig. \ref{Figure3} was implemented. For each patient, we evaluated the average waveform, $\mu_{p(t)}$, (and the standard deviation, $\sigma_{p(t)}$) of the central pressure per beat over the recorded cardiac cycles. Then signal $\mu_{p(t)}$ was compared against the corresponding simulated average waveform, $\mu_{p(t)comp}$. The latter was obtained through the patient-specific multiscale model, which received as model inputs the patient-specific noninvasive data ($S$, $AGE$, $W$, $H$, $HR$, $T_{vc}$, $PP_{b}$ and $P_{m_b}$).

\begin{figure*}[!h]
\begin{center}
\includegraphics*[scale=0.17]{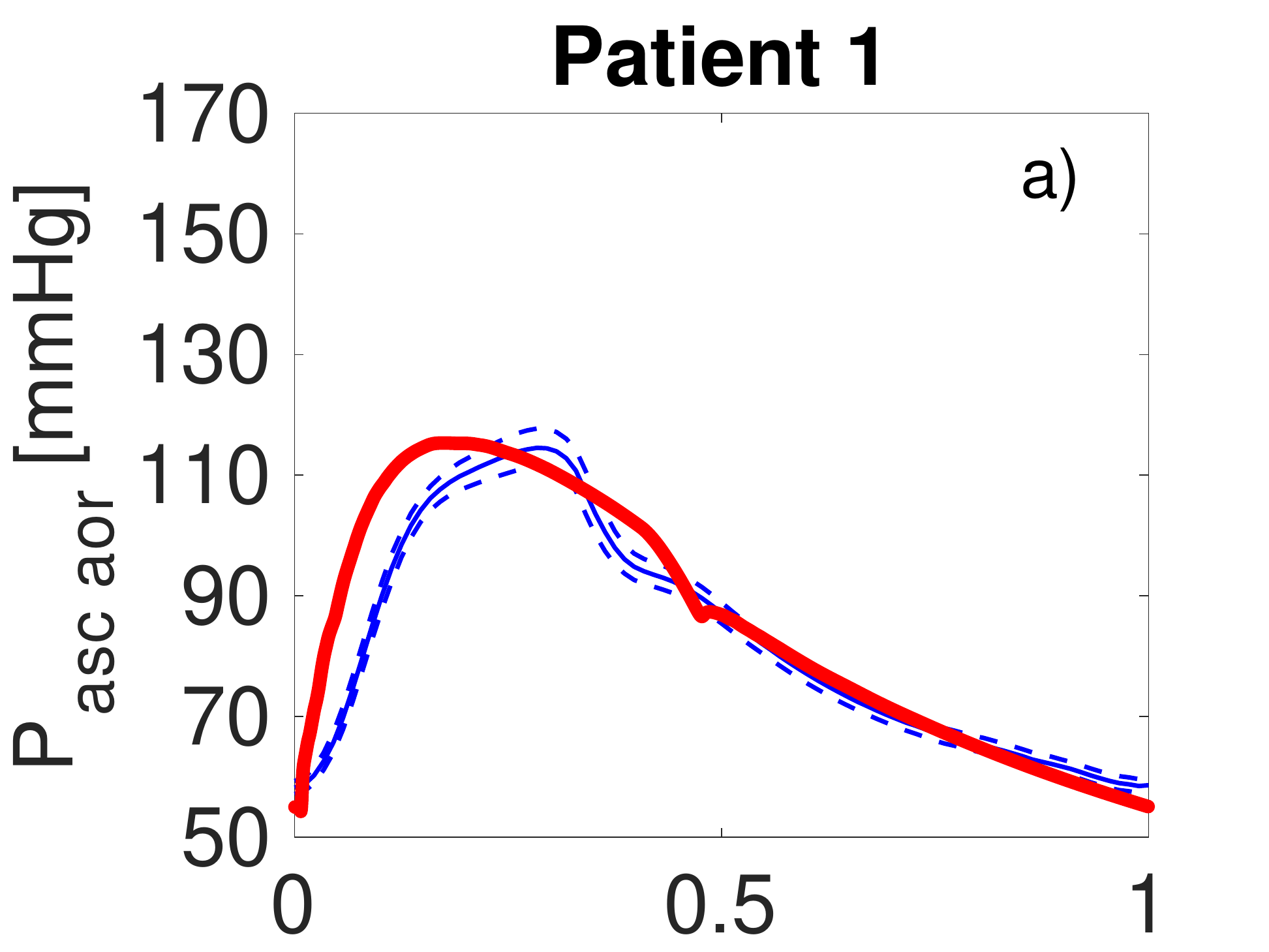}
\includegraphics*[scale=0.17]{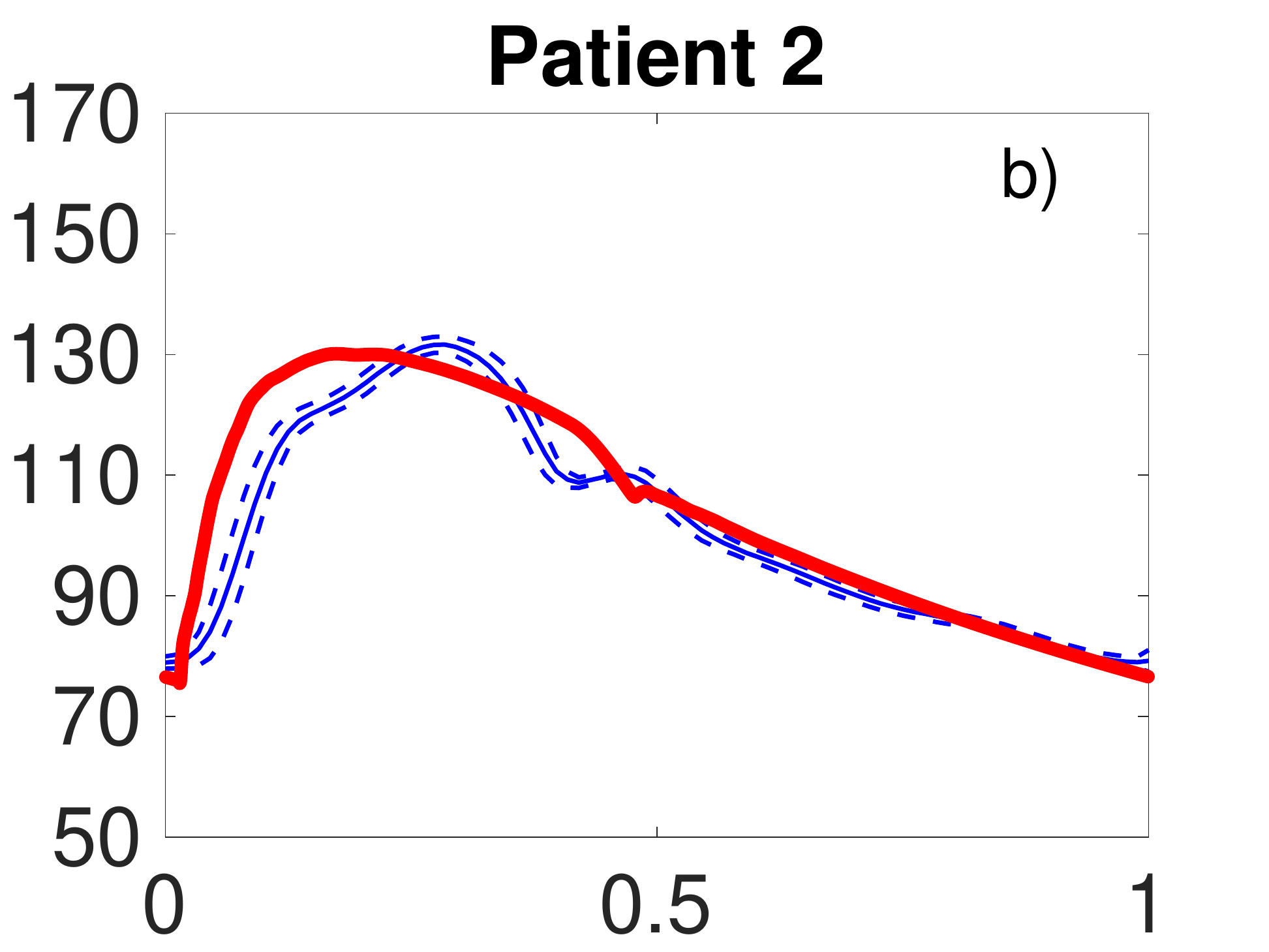}
\includegraphics*[scale=0.17]{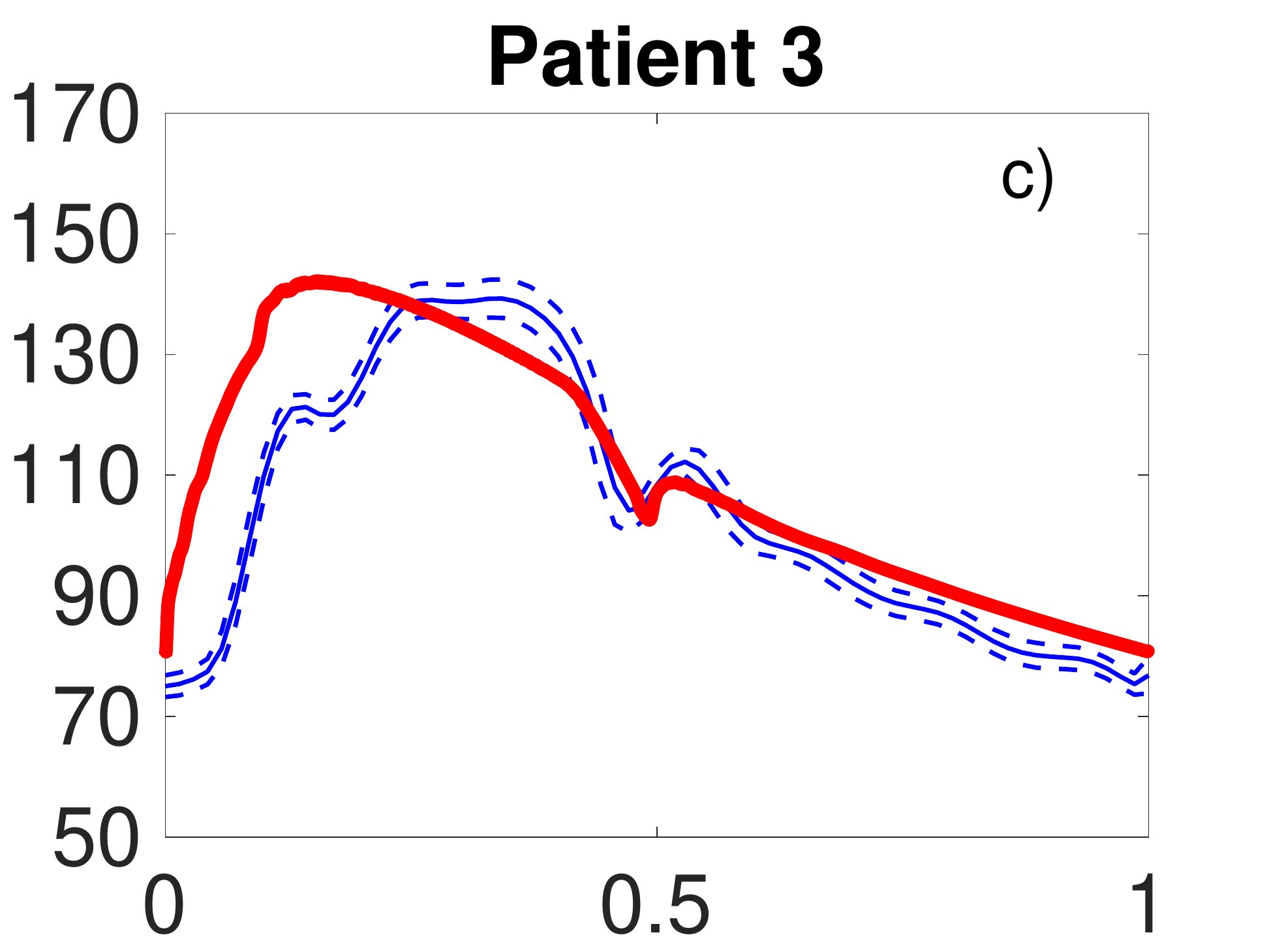}
\includegraphics*[scale=0.17]{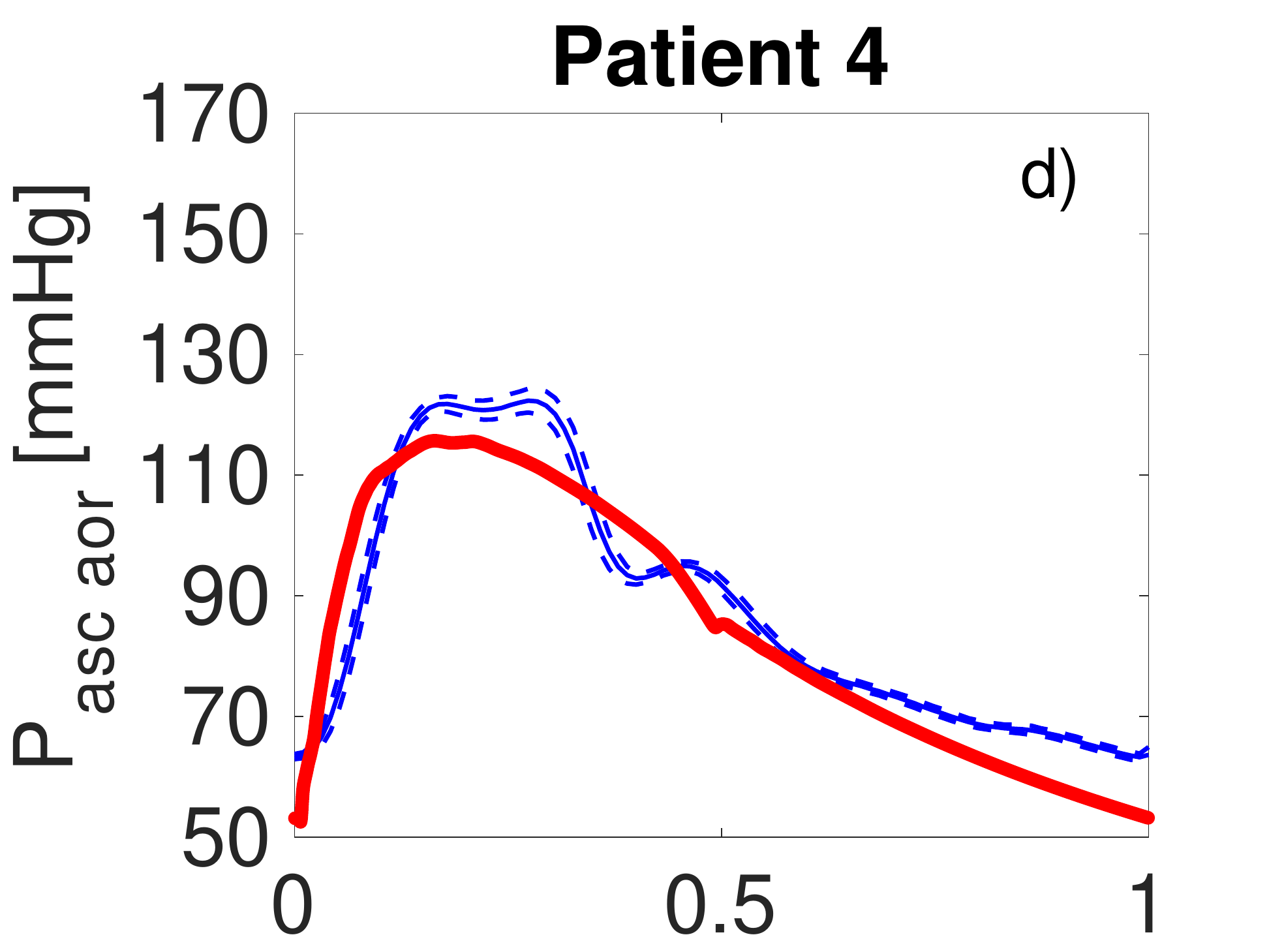}
\includegraphics*[scale=0.17]{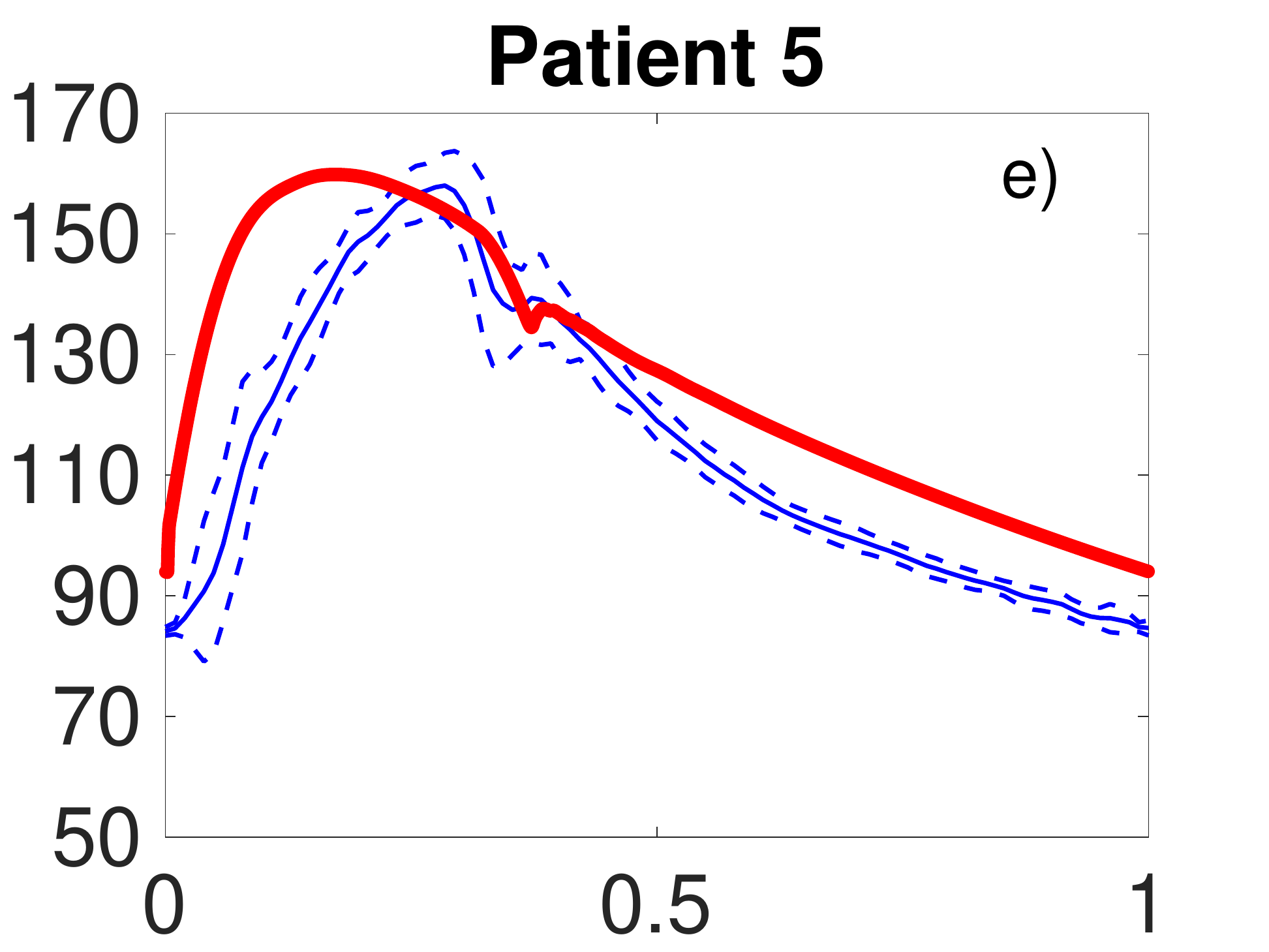}
\includegraphics*[scale=0.17]{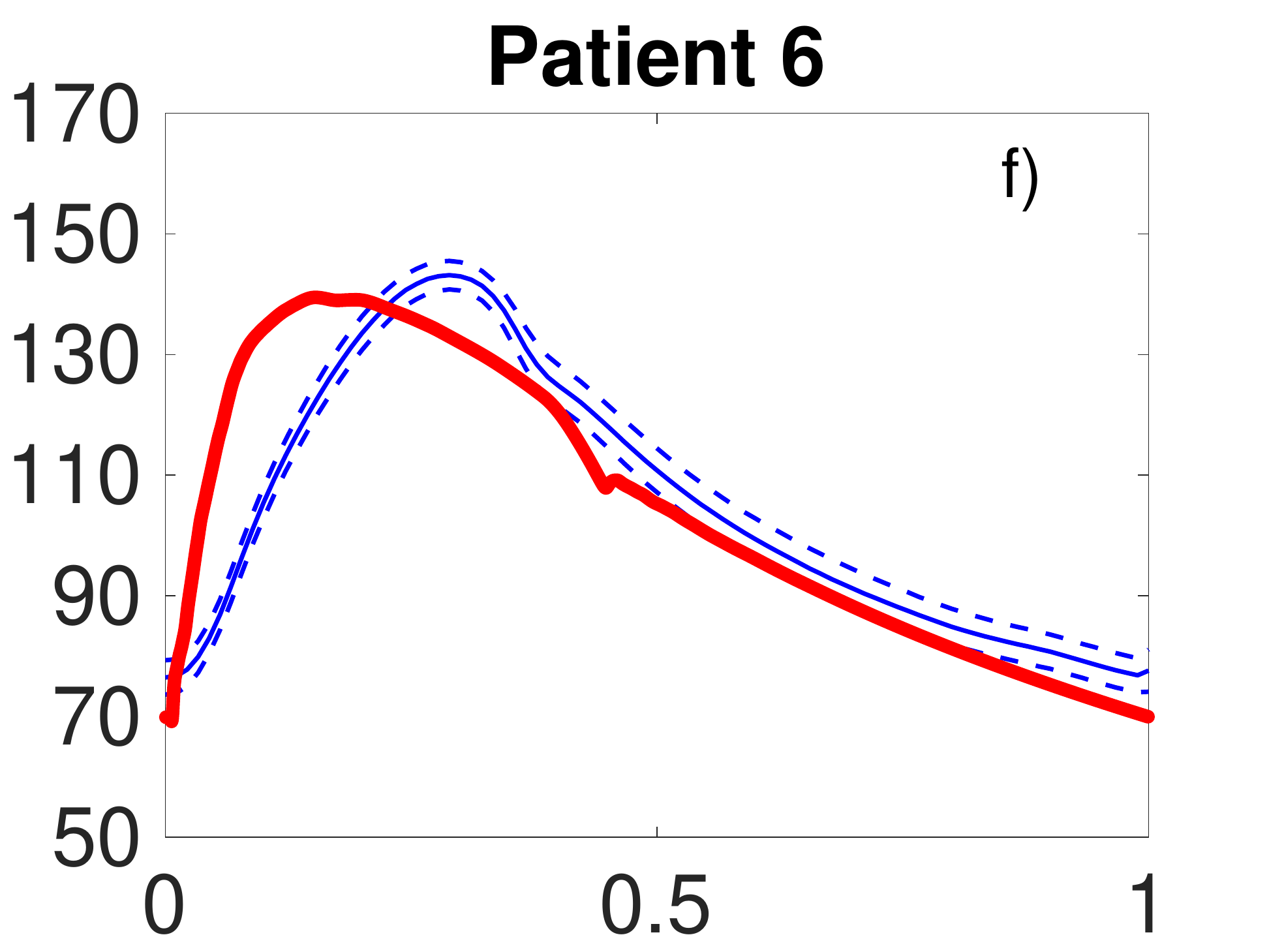}
\includegraphics*[scale=0.17]{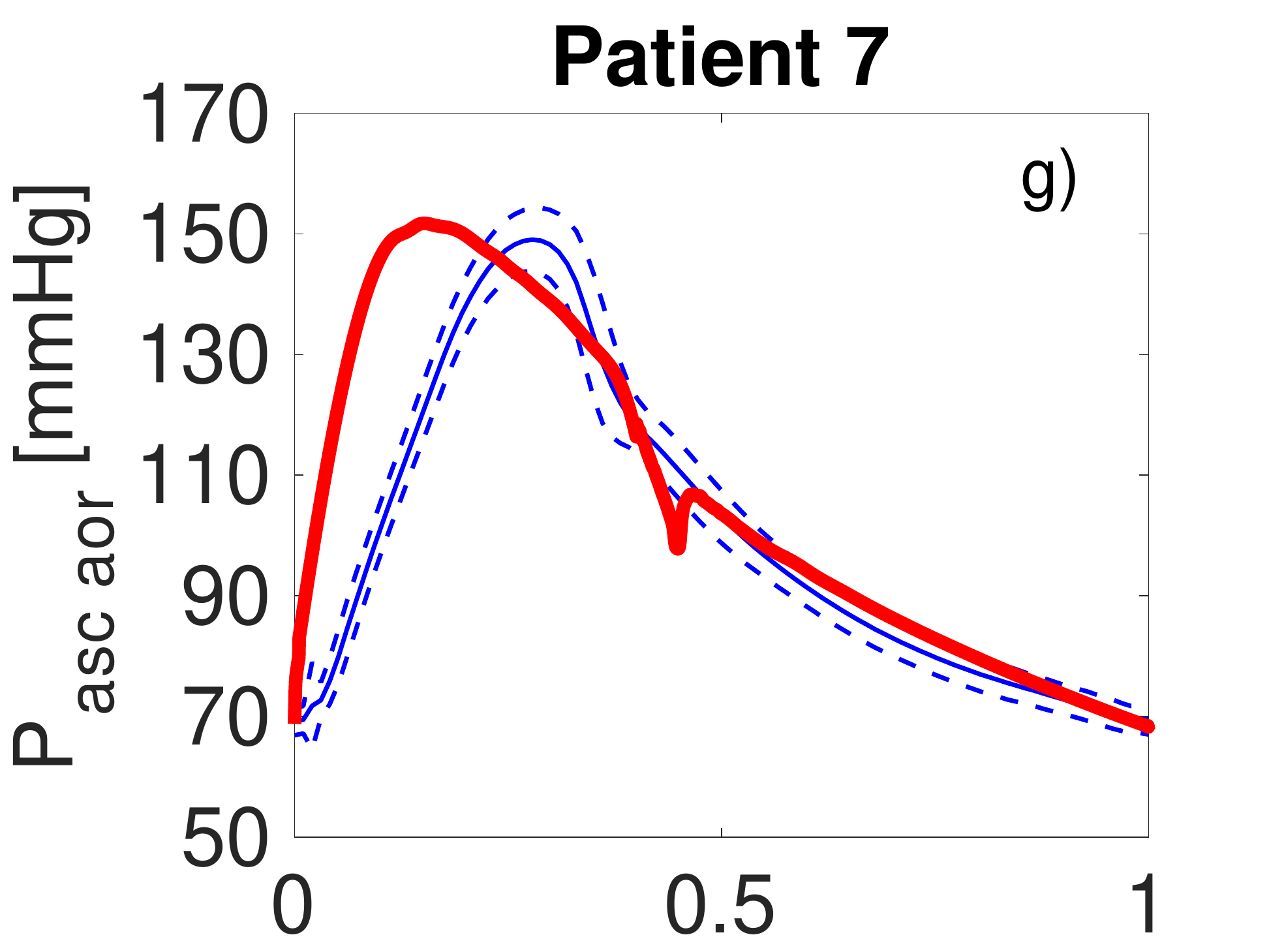}
\includegraphics*[scale=0.17]{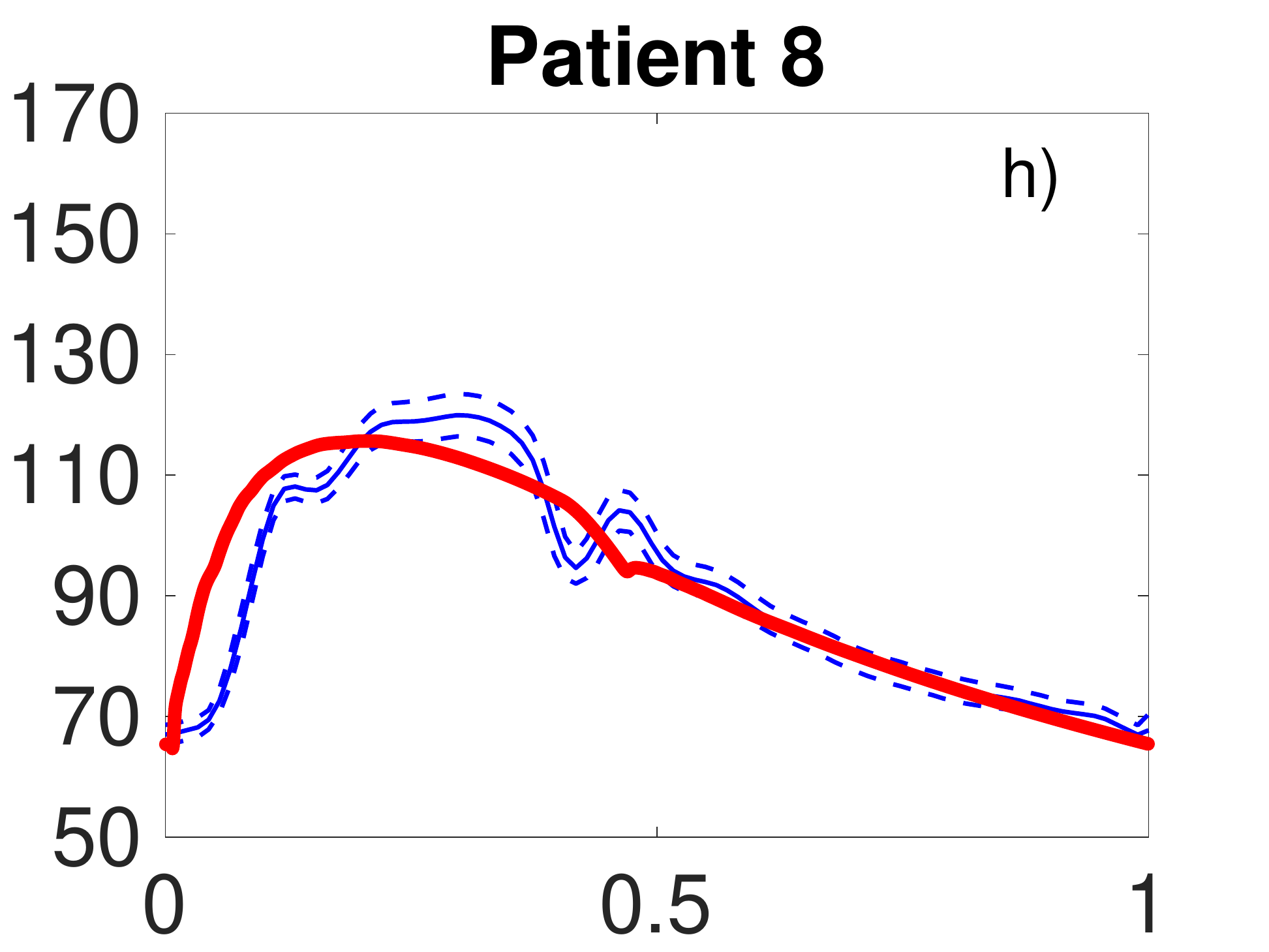}
\includegraphics*[scale=0.17]{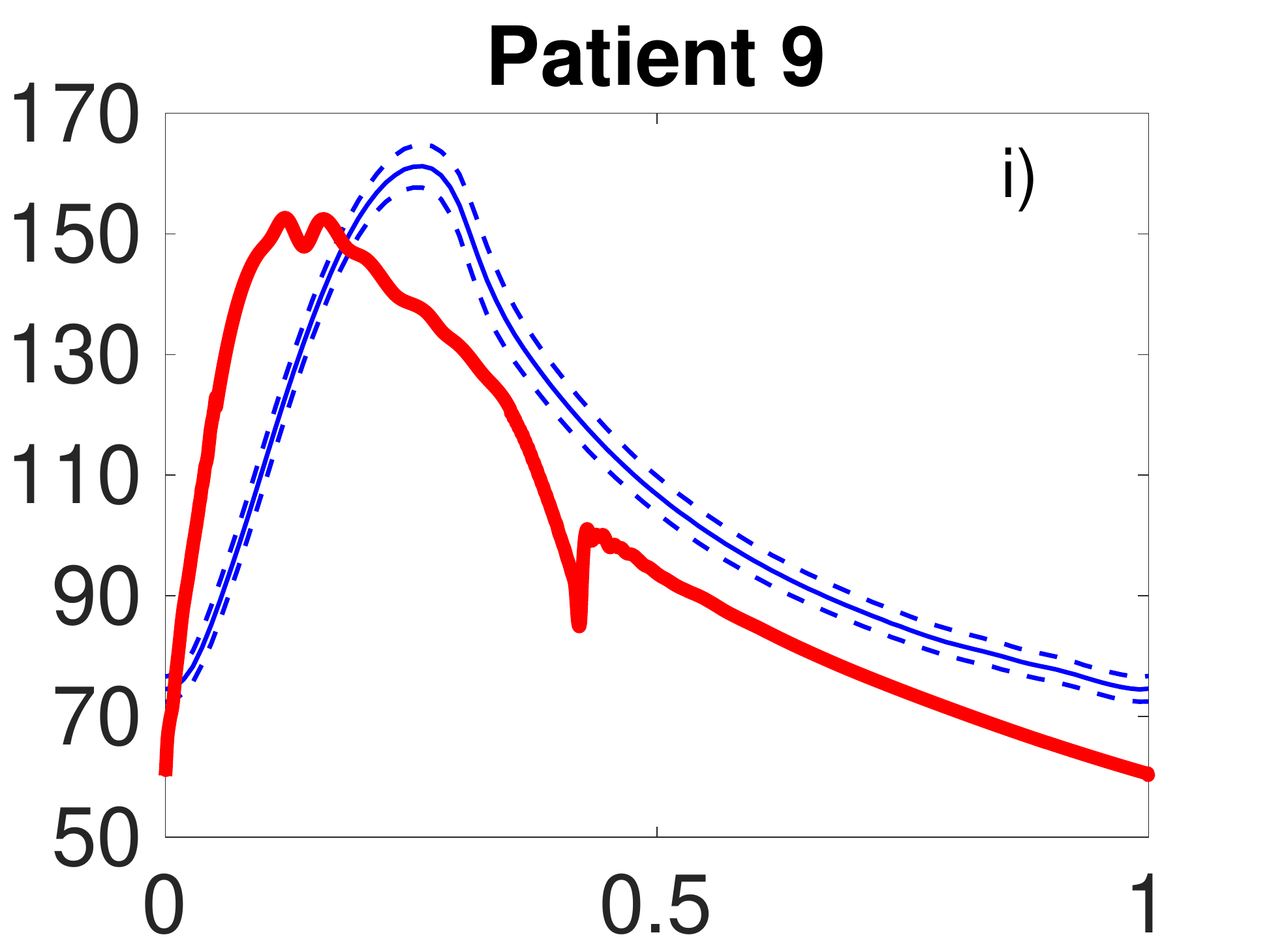}
\includegraphics*[scale=0.17]{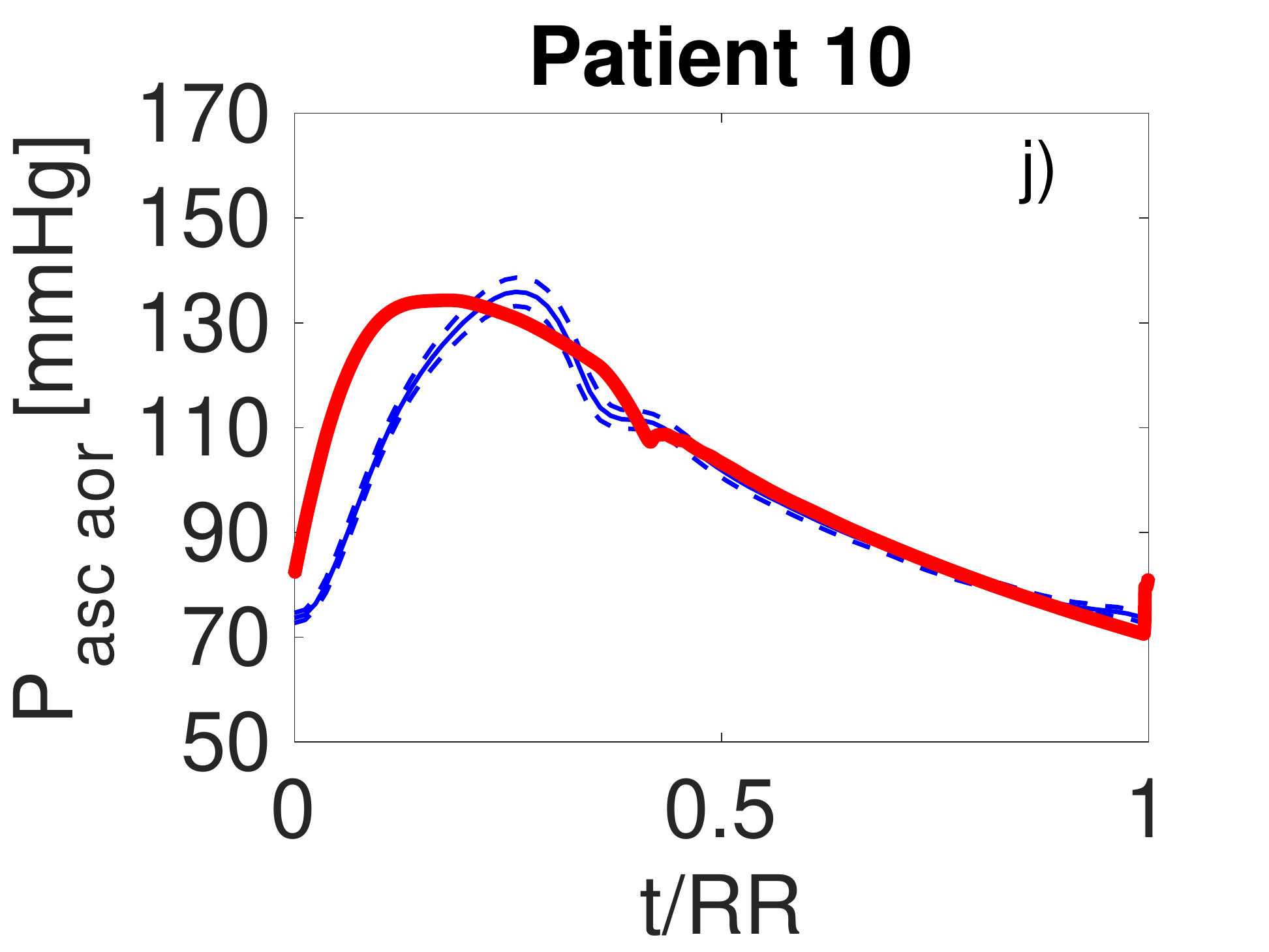}
\includegraphics*[scale=0.17]{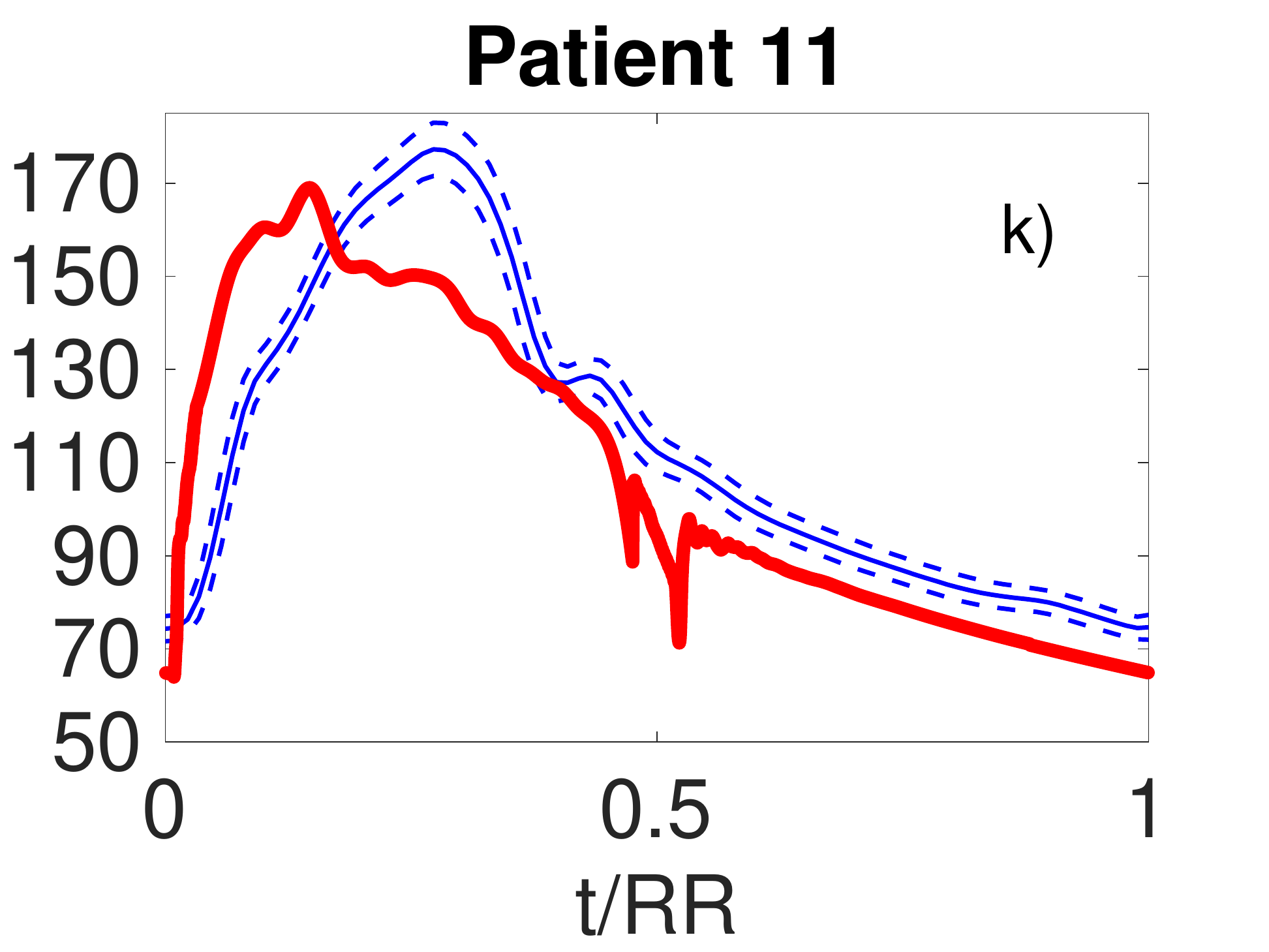}
\includegraphics*[scale=0.17]{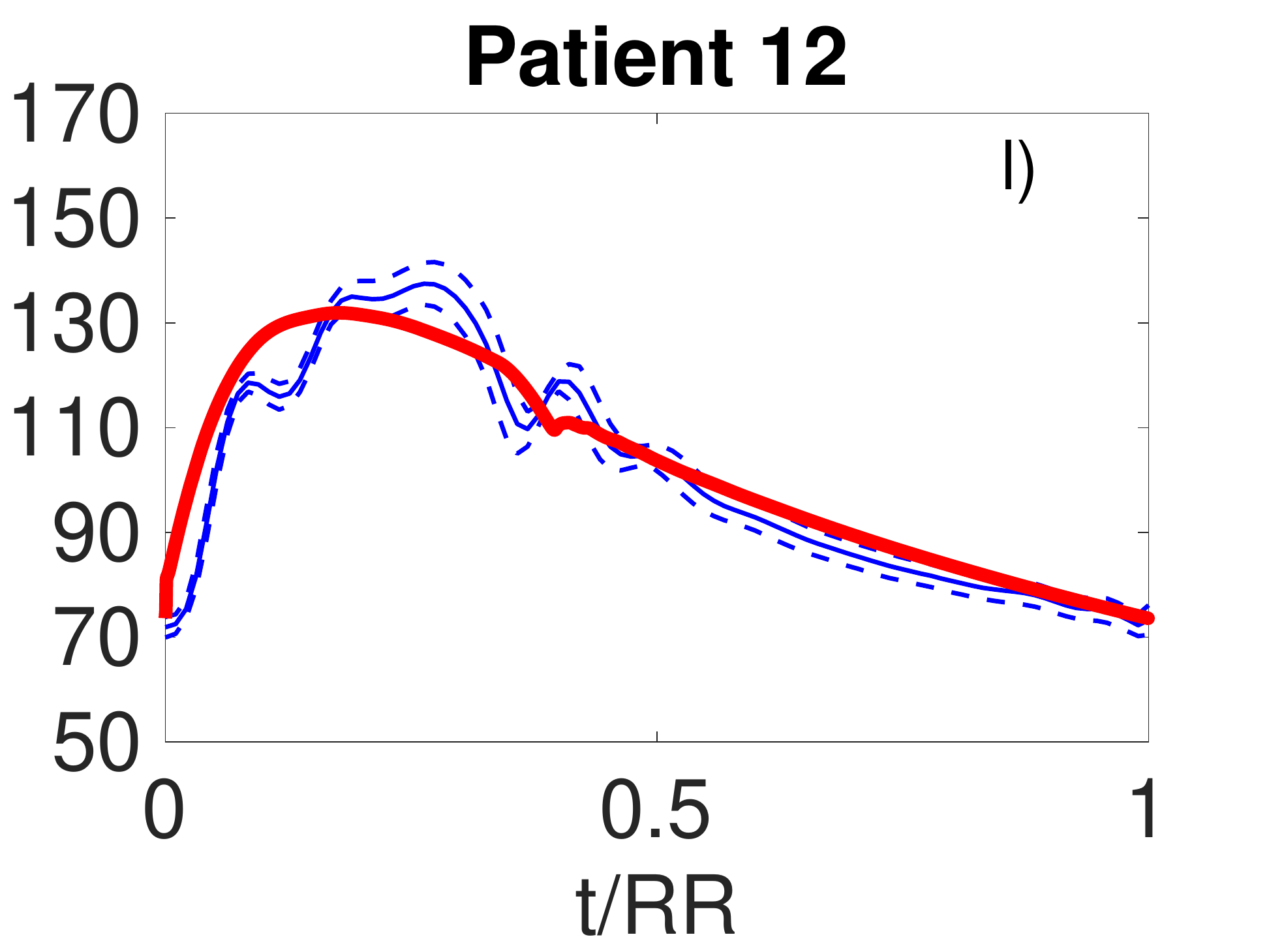}
\caption{}
\label{Figure4}
\end{center}
\end{figure*}

Fig. \ref{Figure4} displays measured central pressure signals ($\mu_{p(t)}$, continuous thin blue line, and $\mu_{p(t)}\pm\sigma_{p(t)}$, dotted thin blue line) and the corresponding simulated average signals ($\mu_{p(t)comp}$, continuous thick red line) for all patients. All signals are reported as a function of the non-dimensional mean heartbeat period, $RR$. Visually, one can appreciate that the simulated signals well match the average measured signals, despite dissimilarities between the measured and computed shapes of the central pressure waveforms.

To better quantify the accuracy of the model, we also determined, for each patient, the errors introduced by both the reference model (the one without the patient-specific adjustments) and the patient-specific model in estimating central systolic, diastolic, mean and pulse pressures with respect to the mean values of the related measured pressures indicated in Table \ref{Tab4}. These results are reported in Table \ref{Tab5}. It emerges that the reference model leads to higher mean errors in central systolic (26.12 mmHg), mean (10.43 mmHg) and pulse (28.46 mmHg) pressures compared with the patient-specific model (4.26, 4.98 and 3.51 mmHg for systolic, mean and pulse pressures, respectively). Differently, slightly smaller mean errors occur in central diastolic pressure (5 mmHg) with respect to the patient-specific model (5.86 mmHg). Based on these results, it is apparent that the adaptations to make the reference model patient-specific are effective in reproducing the patient-specific characteristics. For diastolic pressure, however, the mean error by the reference model is moderately littler than by the patient-specific model. The reference model does not lead to smaller errors in central diastolic BP for all the patients but for only five of the twelve patients, for which the error is drastically reduced. Since these five patients belong to three over the four recognized BP phenotypes (see Table \ref{Tab4}), it is difficult to identify clear correlations between the error in central diastolic BP and one or more BP phenotypes. Thus, it should be verified on a greater number of patients whether (i) central diastolic pressure is smaller with the reference than with the patient-specific model, and (ii) potential correlations between one or more BP phenotypes and the individual errors in central diastolic BP exist.

By adopting the patient-specific model, differences between measured and modeled mean pressures appear quite acceptable. In fact, mean error is always $\leq$5 mmHg, apart from diastolic BP, and standard deviation is always $\leq$8 mmHg, which have been proposed as the minimum acceptable errors in central BP validation protocol \cite{Sharman}.
Coherently with Fig. \ref{Figure4}, Table \ref{Tab5} confirms that diastolic pressure errors are greater than systolic ones for the majority of patients. One can also observe that errors in mean pressure are between errors in systolic and diastolic pressures, while large/small pulse pressure errors do not necessarily correspond to large/small systolic and diastolic pressure errors.

\begin{figure*}[!h]
\begin{center}
\includegraphics*[scale=0.24]{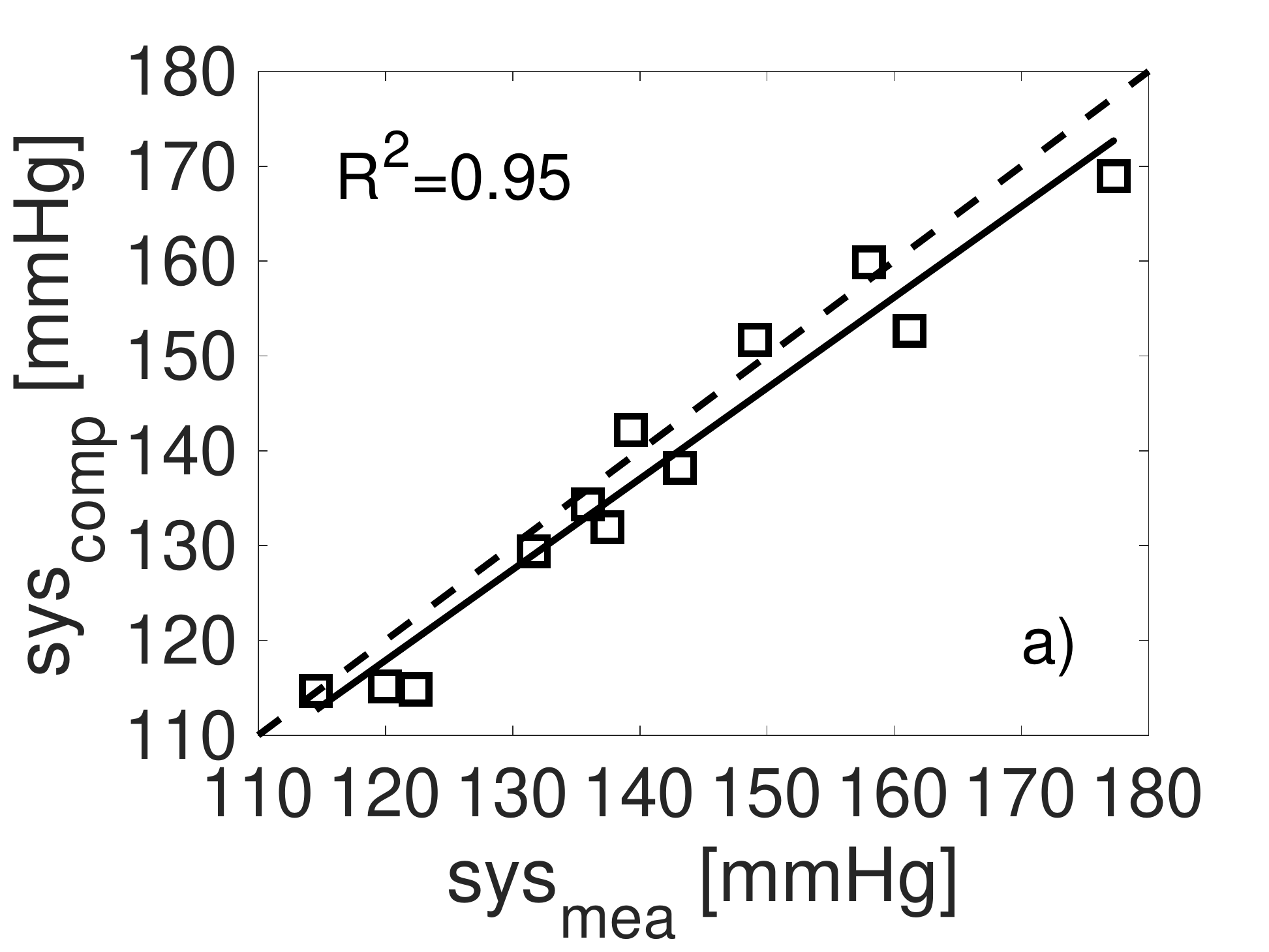}\quad
\includegraphics*[scale=0.24]{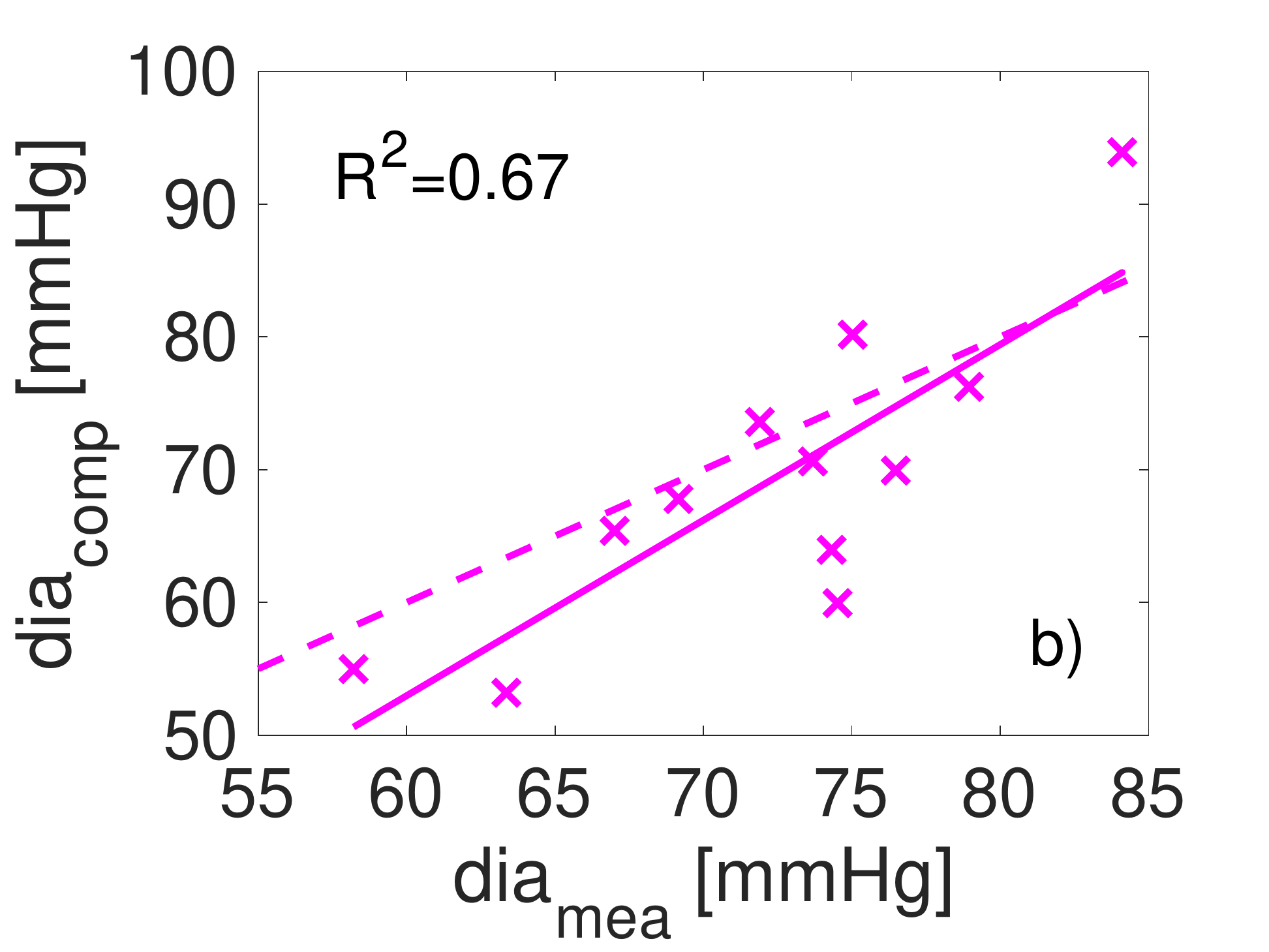}\quad
\caption{}
\label{Figure5}
\end{center}
\end{figure*}

Scatter plots between simulated central systolic/diastolic BPs and catheter measurements are given in Fig. \ref{Figure5}, together with coefficients of determination ($R^2$). The latter are equal to 0.95 and 0.67 for systolic and diastolic pressures, respectively, thereby reflecting a good correlation between measured and simulated pressures, as well as larger errors by the model for diastolic pressures than for systolic pressures.

\begin{figure*}[!h]
\begin{center}
\includegraphics*[scale=0.24]{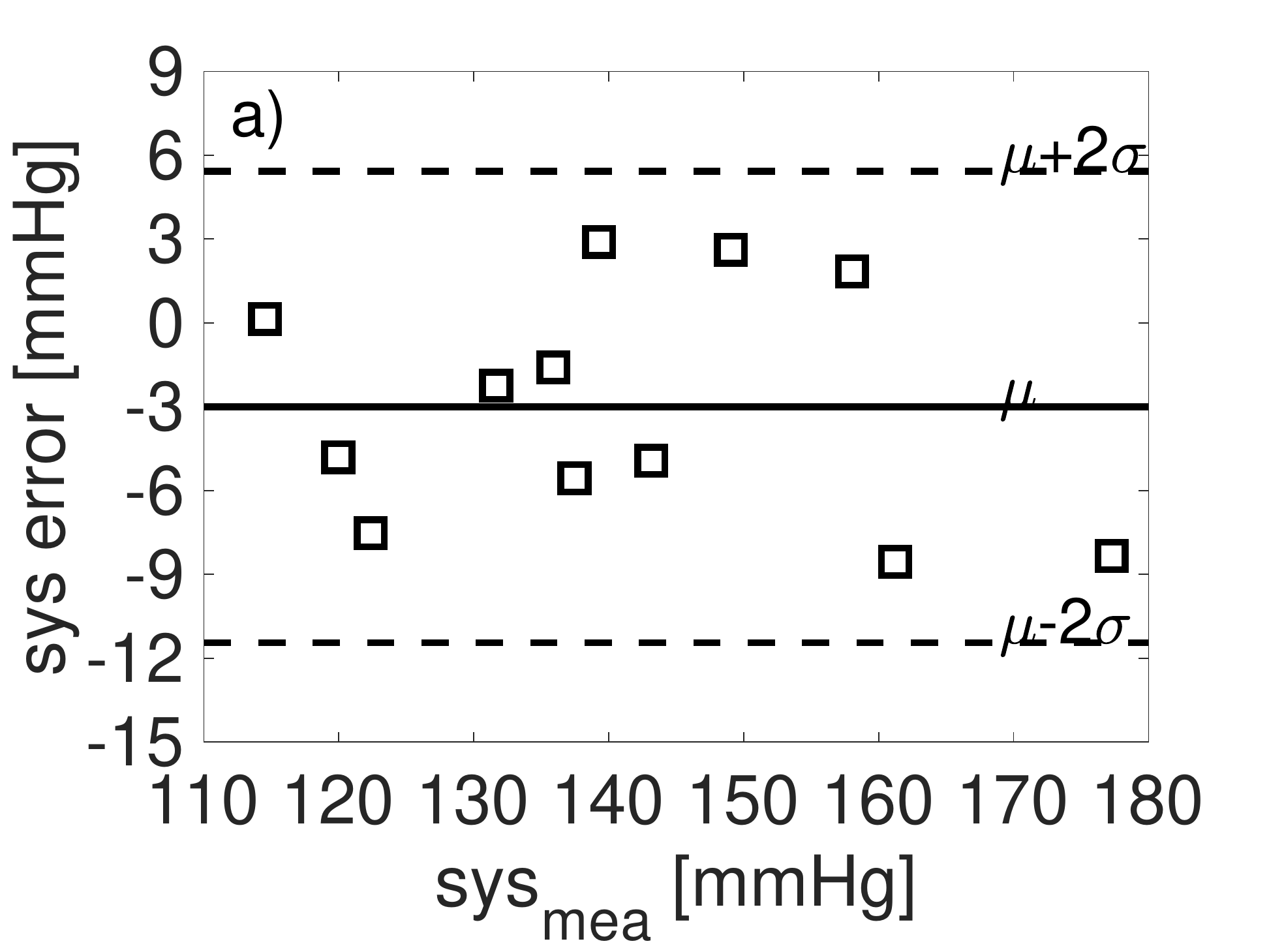}\quad
\includegraphics*[scale=0.24]{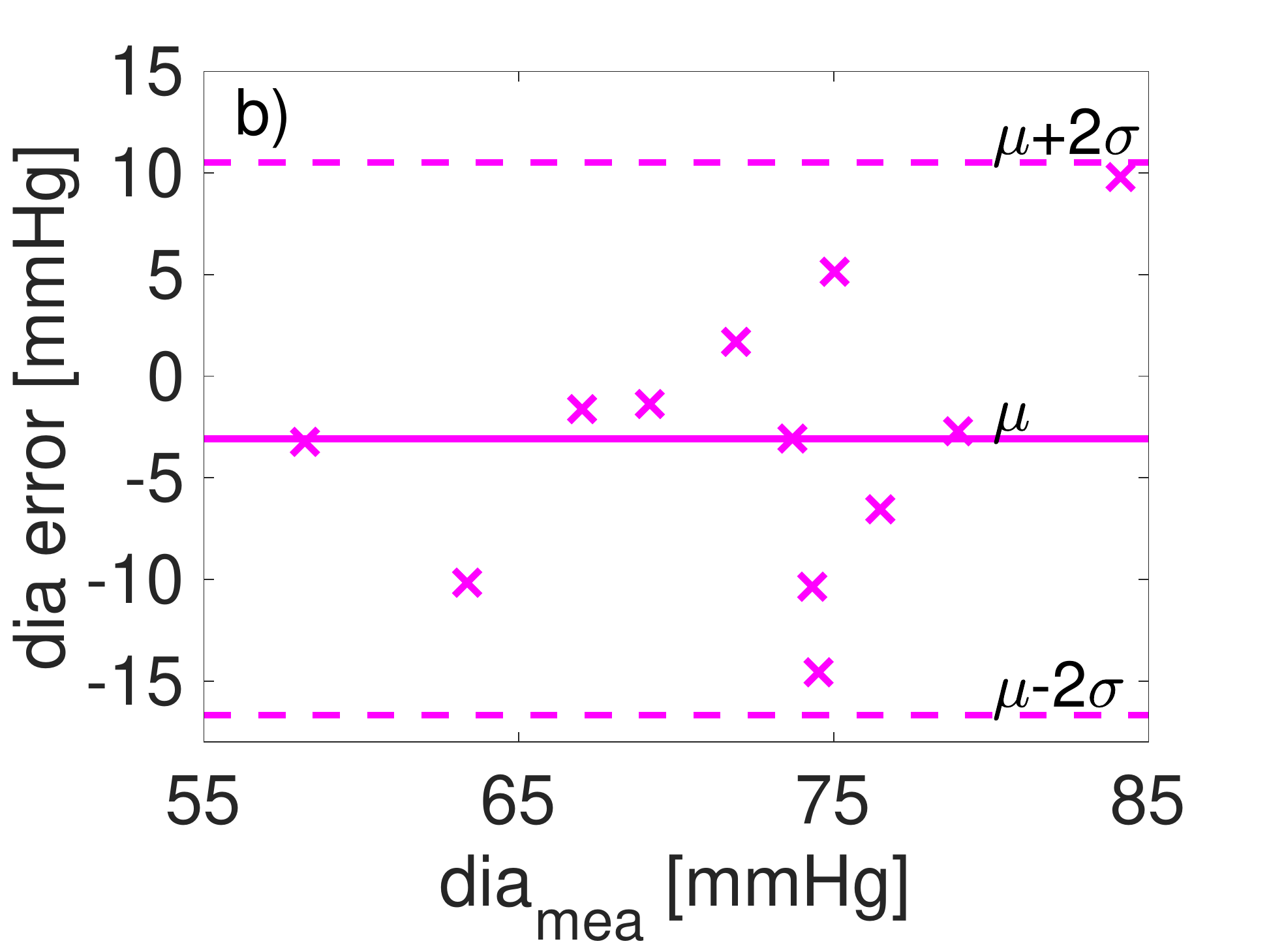}\quad
\caption{}
\label{Figure6}
\end{center}
\end{figure*}

Bland-Altman plots of the central systolic and diastolic pressure errors by the model are depicted in Fig. \ref{Figure6}. The coefficients of determination, $R^2$, for the relations between the central pressure error and the mean central pressure are 0.035 for the systolic pressure and 0.11 for the diastolic pressure. Thus, the present data suggest that the error magnitude does not depend on the systolic pressure, even if it is slightly correlated to the diastolic pressure.

\section{Discussion}

In the present study, we propose a patient-specific mathematical model to noninvasively evaluate central BP, and test it against the invasive central BP measurements of twelve patients.

Taking into account that this computational method does not require invasive pressure signals as input data, and it is just based on systolic/diastolic automatic oscillometric brachial pressures and generic anthropometric parameters, the comparison between computed and measured central pressure waveforms seems satisfying. More in details, errors (absolute mean errors $\pm$ standard deviation) in systolic, diastolic, mean and pulse pressure for the twelve patients are 4.26$\pm$2.81 mmHg, 5.86$\pm$4.38 mmHg, 4.98$\pm$3.95 mmHg and 3.51$\pm$2.38 mmHg, respectively.

In order to contextualize the approach we propose within the current medical scenario, it is useful to compare the errors obtained through our method with those produced by one of the most common device to noninvasively estimate central BP, the SphygmoCor, and generalized/variable transfer functions. The SphygmoCor leads to errors in central systolic, diastolic and pulse pressure of -8.2$\pm$10.3 mmHg, 7.6$\pm$8.7 mmHg and -12.2$\pm$10.4 mmHg, respectively, with the calibration performed through brachial cuff pressure. These errors were obtained from a large meta-analysis including 857 subjects \cite{Cheng}. Comparing these results with the ones presented in this study, our method seems promising, although it still needs to be tested in an extended cohort of individuals. In other works by Shih et al. \cite{Shih1,Shih2}, where central BP was evaluated through an ensemble-averaged generalized transfer function extracted from 40 individual transfer functions, smaller mean errors (in modulus) in central systolic BP (-2.2$\pm$6.4 mmHg in \cite{Shih1} and -2.1$\pm$7.7 mmHg in \cite{Shih2}) were found, albeit larger errors in central pulse BP emerged (10.3$\pm$8 mmHg in \cite{Shih1}) with respect to the ones by our patient-specific model. More recently, a new physiology-based technique was published by Natarajan et al. \cite{Natarajan} to accurately derive central BP via a standard automatic arm cuff. In particular, they applied a variable transfer function method to the brachial BP-like waveform, which was in turn derived through an ensemble averaging/calibration procedure, with brachial systolic and diastolic BPs for calibration obtained from the application of a patient-specific method to an oscillogram. The technique by Natarajan et al. yields errors in central systolic, diastolic and pulse BP between -0.6 and 2.6 mmHg and corresponding standard deviation in the range 6.8-9 mmHg. Thus, despite the proposal by \cite{Natarajan} produces extremely reduced mean errors with respect to our model, standard deviations double compared to ours.
Based on this comparison, our model produces competitive results with respect to the SphygmoCor and the generalized transfer function by Shih et al. \cite{Shih1,Shih2}, despite leading to larger mean errors than the physiology-based technique by Natarajan et al. \cite{Natarajan}. However, considering a wider picture of medical devices to estimate central BP, the modeling approach here proposed gives comparable results. In fact, according to Papaioannou et al. \cite{Papaioannou} - which took into account 22 validation studies of 11 medical devices involving a total of 808 subjects - the error in aortic systolic BP through noninvasive brachial BP calibration is between -7.79 and -3.84 mmHg. Our patient-specific model, which leads to a mean error in central systolic BP of -4.26 mmHg, results among the best 11\% of the methods considered by \cite{Papaioannou} and adopting noninvasively measured brachial BP values for calibration. Even if this study is not able to provide a definitive answer about the best prognostic parameter between central and brachial BP, it offers a promising preliminary view of the \textit{in-silico} approach in providing patient-specific central BP estimations, and represents a stimulus to exploit similar models to estimate additional individual hemodynamic measures. Patient-specific models, in fact, apart for central BP evaluation, can be exploited to obtain further cardiac and vascular parameters, which can enrich the hemodynamic picture of each patient.

It is clear that the errors by our modeling approach largely depend on the errors associated to the brachial oscillometric measurement, from which the brachial mean and pulse BPs ($P_{m_b}$ and $PP_b$) adopted in the patient-specific calibration of the reference model were extracted. The importance of these two parameters in making the reference model patient-specific is, in fact, proved by the sensitivity analysis reported in section \ref{sensitivity_analysis}. Considering the latter, it is predictable that any error associated to the input data $P_{m_b}$ and $PP_b$ is transmitted (with the same order of magnitude) to the output data, namely the central BP values. Despite the automatic oscillometric device we adopted to assess brachial systolic and diastolic BP is widely used in clinical practice, no peer-reviewed clinical validation information is available on this technology. It follows that a direct comparison between the input errors in the values of $P_{m_b}$ and $PP_b$ and the output errors in the estimations of central BP for the 12 patients is not possible here. Together with the brachial oscillometric measurements, also invasive BP recordings could be affected by bias. In fact, we did not perform any specific assessment of the frequency response of our pressure measuring system, possibly introducing bias to our invasive BP measurements. As previously mentioned, no information on the wave form sampling rate of the pressure signal recording system has been provided by the manufacturer, that represents another potential source of bias in invasive BP values.

As next steps in successive studies, we first recognize the need to increase the number of test patients, which would allow one to estimate the actual model errors in central BP, and identify potential correlations between central pressure errors and impacting factors, in order to further reduce the model errors. Then, a direct comparison between input and output errors should be executed, and output errors could be possibly reduced through more precise patient-specific oscillometric BP measurement methods, like the one proposed by Liu et al. \cite{Liu} and adopted by Natarajan et al. \cite{Natarajan}.

\section{Conclusions}

The proposed modeling approach exhibits a good patient-specific response. Despite it is not demonstrated to be superior to all the other methodologies to noninvasively estimate central BP, it shows acceptable approximation levels for central systolic BP evaluation with noninvasive brachial BP calibration, and could be adopted not only for central BP but also for other central cardiac and vascular hemodynamic parameters on specific subjects. Further efforts are needed to test the reliability of the present method, by extending the whole procedure to a larger cohort of individuals, and possibly reducing the input errors through more accurate oscillometric BP measurements techniques.

%\begin{acknowledgements}
%If you'd like to thank anyone, place your comments here
%and remove the percent signs.
%\end{acknowledgements}

% Authors must disclose all relationships or interests that
% could have direct or potential influence or impart bias on
% the work:
%
% \section*{Conflict of interest}
%
% The authors declare that they have no conflict of interest.

% BibTeX users please use one of
%\bibliographystyle{spbasic}      % basic style, author-year citations
%\bibliographystyle{spmpsci}      % mathematics and physical sciences
%\bibliographystyle{spphys}       % APS-like style for physics
%\bibliography{}   % name your BibTeX data base

\section*{Declarations}

\section*{Funding}
Not applicable.

\section*{Conflict of interest/Competing interests}
The authors declare no competing interests.

\section*{Availability of data and material/Code availability}
Data and material/code produced and analyzed during this study are available on reasonable request.

\section*{}
\textit{All procedures followed were in accordance with the ethical standards of the responsible committee on human experimentation (institutional and national) and with the Helsinki Declaration of 1975, as revised in 2000 (5). Informed consent was obtained from all patients for being included in the study. No animal studies were carried out by the authors for this article.}

\newpage

% Non-BibTeX users please use

\section*{Figure legends}

\textbf{Fig. 1} Schematic representation of the procedure adopted to make the reference mathematical model patient-specific. $HR$, $T_{vc}$ and $P_{m_b}$ are directly introduced to the model, while $H$, $BSA$ (which is function of $H$ and $W$ through the Du Bois's formula \cite{Du_Bois}), $AGE$, $S$ and $PP_b$ are used in the empirical relations to adapt the model parameters - arterial lengths, $L_{art}$, aortic/carotid arterial diameters, $D_{ca}$, aortic/carotid arterial thicknesses, $h_{ca}$, aortic/carotid pulse wave velocities, $PWV_{ca}$, end-systolic/end-diastolic left-ventricular elastance values, $E_{es/ed}$, and venous compliances, $C_{v}$ - to the patient characteristics, starting from the reference values (subscript $ref$)

\textbf{Fig. 2} Sensitivities of central systolic (sys), diastolic (dia), mean (mean) and pulse (pp) pressures to the input model parameters (age, $AGE$, weight, $W$, height, $H$, heart rate, $HR$, mean left ventricular contraction time, $T_{vc}$, and mean and pulse brachial BPs, $P_{m_b}$ and $PP_b$) for men

\textbf{Fig. 3} Schematic representation of the procedure implemented to test the patient-specific multiscale model

\textbf{Fig. 4} Measured central pressure signals ($\mu_{p(t)}$, continuous thin blue line, and $\mu_{p(t)}\pm\sigma_{p(t)}$, dotted thin blue line) and corresponding simulated average signals ($\mu_{p(t),comp}$, continuous thick red line). Time is made dimensionless by the heartbeat period, $RR$

\textbf{Fig. 5} Scatter plots between computed (subscript comp) systolic (sys) and diastolic (dia) pressures and measured (subscript mea) values. $R^2$ is the coefficient of determination and continuous/dotted lines are the those of linear regression/equality

\textbf{Fig. 6} Bland-Altman plots of the central systolic (sys) and diastolic (dia) pressures errors by the model (sys err and dia err, respectively). $\mu$ and $\sigma$ stand for the mean and standard deviation values of the pressure errors

\section*{Figures}

\clearpage

\section*{Tables}

\begin{table}[ht!]
\caption{Regression coefficients expressing the variations in aortic lengths ($c_1$) - for both women (subscript $_{,w}$) and men (subscript $_{,m}$) - and pulse wave velocities ($c_5$ linear coefficients and $c_6$ quadratic coefficients) with $AGE$ according to Rylski et al. \cite{Rylski} and Hickson et al. \cite{Hickson}, respectively. Results are given for 4 different aortic tracts; 1: from the aortic valve to the origin of the brachiocephalic artery, 2: from the end of tract 1 to the the origin of the left subclavian artery, 3: from the end of tract 2 to the origin of coeliac artery, and 4: from the end of tract 3 to the aortic bifurcation.}
\begin{center}
\begin{scriptsize}
\begin{tabular}{c@{\quad}c@{\quad}c@{\quad}c@{\quad}c}
\hline\noalign{\smallskip}
\rule{0pt}{12pt}
Aortic tract&$c_{1,w}$&$c_{1,m}$&$c_{5}$&$c_{6}$\\
&[mm/y/m$^2$]&[mm/y/m$^2$]&[mm/y]&[mm/y$^2$]\\
\noalign{\smallskip}\hline\noalign{\smallskip}
\rule{0pt}{12pt}
1 & 0.22 & 0.21 & 1.8986 10$^{-4}$ & 0.0266 \\
2 & 0.11 & 0.09 & 8.1551 10$^{-4}$ & 0.0016 \\
3 & 0.45 & 0.55 & 6.3056 10$^{-4}$ & 0.0109 \\
4 & 0.19 & 0.19 & 5.0331 10$^{-4}$ & 0.0674 \\
\noalign{\smallskip}\hline
\end{tabular}
\end{scriptsize}
\end{center}
\label{Tab1}
\end{table}

\begin{table}[ht!]
\caption{Regression coefficients for women ($c_{2,w}/c_{3,w}$) and men ($c_{2,m}/c_{3,m}$) expressing the variations in aortic and carotid diameters with $AGE$/$BSA$. Dependencies with $AGE$ are formulated according to Rylski et al. \cite{Rylski} (in [mm/y/m$^2$]) and Kamenskiy et al. \cite{Kamenskiy} (in [mm/y]) for the aortic and carotid tracts, while dependencies with $BSA$ according to Davis et al. \cite{Davis} and Krejza et al. \cite{Krejza} (both in [mm/m$^2$]) for the aortic and carotid tracts, respectively. Results are given for 6 different arterial tracts; 1: from the aortic valve to the origin of the brachiocephalic artery, 2: from the end of tract 1 to the the origin of the left subclavian artery, 3: from the end of tract 2 to the origin of coeliac artery, 4: from the end of tract 3 to the aortic bifurcation, 5: along the right common carotid artery, and 6: along the left common carotid artery.}
\begin{center}
\begin{scriptsize}
\begin{tabular}{c@{\quad}c@{\quad}c}
\hline\noalign{\smallskip}
\rule{0pt}{12pt}
Arterial tract&$c_{2w}, c_{3w}$&$c_{2,m}, c_{3m}$\\
\noalign{\smallskip}\hline\noalign{\smallskip}
\rule{0pt}{12pt}
1 & 0.11, 5.6 &  0.08, 7.6 \\
2 & 0.08, 2.8 &  0.06, 5.5 \\
3 & 0.08, 3.1 &  0.08, 3.8 \\
4 & 0.05, 3.1 &  0.06, 3.8 \\
5 & 0.017, 1.13 & 0.018, 1.21 \\
6 & 0.006, 1.13 & 0.006, 1.21 \\
\noalign{\smallskip}\hline
\end{tabular}
\end{scriptsize}
\end{center}
\label{Tab2}
\end{table}

\begin{table}[ht!]
\caption{Anthropometric and clinical data, presence of comorbidities, like diabetes (d) and ischemic heart disease (IHD), and smoking status (s) of patients. Averages and standard deviation values of $AGE$, $W$, $H$, $HR$, $T_{vc}$, $PP_{b}$ and $P_{m_b}$ are reported in bold in the last row.}
\begin{center}
\begin{scriptsize}
\begin{tabular}{c@{\quad}c@{\quad}c@{\quad}c@{\quad}c@{\quad}c@{\quad}c@{\quad}c@{\quad}c@{\quad}c@{\quad}c@{\quad}c}
\hline\noalign{\smallskip}
\rule{0pt}{12pt}
Patient&$S$&$AGE$&$W$&$H$&$HR$&$T_{vc}$&$PP_{b}$&$P_{m_b}$&d&IHD&s\\
number & & [years] & [kg] & [cm] & [bpm] & [s] & [mmHg] & [mmHg]&&&\\[2pt]
\noalign{\smallskip}\hline\noalign{\smallskip}
\rule{0pt}{12pt}
1 & m & 72 & 61 & 170 & 63 & 0.38 & 63 & 75 & x & x &\\
2 & f & 68 & 81 & 175 & 65 & 0.39 & 50 & 94 & & x & x\\
3 & m & 73 & 81 & 170 & 80 & 0.34 & 62 & 101 & & &\\
4 & f & 83 & 68 & 163 & 61 & 0.39 & 63 & 74 & & &\\
5 & m & 65 & 99 & 193 & 54 & 0.39 & 83 & 116 & & & x\\
6 & f & 81 & 69 & 167 & 63 & 0.36 & 74 & 93 & x & x & x\\
7 & m & 75 & 91 & 172 & 55 & 0.41 & 82 & 96 & x & x & x\\
8 & m & 62 & 77 & 175 & 61 & 0.41 & 55 & 82 & & &\\
9 & f & 72 & 60 & 167 & 51 & 0.39 & 78 & 91 & & & x\\
10 & m & 74 & 97 & 182 & 71 & 0.32 & 60 & 92 & x & &\\
11 & f & 73 & 64 & 162 & 62 & 0.39 & 88 & 99 & & &\\
12 & m & 62 & 82 & 179 & 61 & 0.36 & 61 & 93 & & &\\
& & \textbf{71.67} & \textbf{77.50} & \textbf{172.92} & \textbf{62.25} & \textbf{0.38} & \textbf{68.25} & \textbf{92.17} & & &\\
& & \textbf{$\pm$6.61} & \textbf{$\pm$13.46} & \textbf{$\pm$8.72} & \textbf{$\pm$7.70} & \textbf{$\pm$0.027}& \textbf{$\pm$12.24} & \textbf{$\pm$11.45} & & &\\[2pt]
\noalign{\smallskip}\hline
\end{tabular}
\end{scriptsize}
\end{center}
\label{Tab3}
\end{table}

\begin{table}[ht!]
\caption{Mean and standard deviations values of systolic (sys) and diastolic (dia) invasive pressures along the ascending aorta (AA), right brachial (RBA) and radial (RRA) arteries. Coefficients of variation are provided in percentage. The phenotype associated to each patient (I, II, III, or IV, see text), identified as indicated by Picone et al. \cite{Picone}, is reported in brackets below the Patient number.}
\begin{center}
\begin{scriptsize}
\begin{tabular}{c@{\quad}c@{\quad}c@{\quad}c@{\quad}c@{\quad}c@{\quad}c}
\hline\noalign{\smallskip}
\multirow{3}{*}{Patient number}&\multicolumn{2}{c}{AA}& \multicolumn{2}{c}{RBA}&\multicolumn{2}{c}{RRA}\\
\cline{2-7}
&\multicolumn{1}{c}{sys}&\multicolumn{1}{c}{dia}&\multicolumn{1}{c}{sys}&\multicolumn{1}{c}{dia}&\multicolumn{1}{c}{sys}&\multicolumn{1}{c}{dia}\\
&\multicolumn{1}{c}{[mmHg]}&\multicolumn{1}{c}{[mmHg]}&\multicolumn{1}{c}{[mmHg]}&\multicolumn{1}{c}{[mmHg]}&\multicolumn{1}{c}{[mmHg]}&\multicolumn{1}{c}{[mmHg]}\\
\noalign{\smallskip}\hline\noalign{\smallskip}
\multicolumn{1}{c}{1}&\multicolumn{1}{c}{114.54}&\multicolumn{1}{c}{58.21}&\multicolumn{1}{c}{124.38}&\multicolumn{1}{c}{59.53}&\multicolumn{1}{c}{154.59}&\multicolumn{1}{c}{64.41}\\
\multicolumn{1}{c}{(I)}&\multicolumn{1}{c}{$\pm$3.21}&\multicolumn{1}{c}{$\pm$1.10}&\multicolumn{1}{c}{$\pm$4.07}&\multicolumn{1}{c}{$\pm$1.81}&\multicolumn{1}{c}{$\pm$2.33}&\multicolumn{1}{c}{$\pm$1.50}\\
\multicolumn{1}{c}{}&\multicolumn{1}{c}{2.80\%}&\multicolumn{1}{c}{1.89\%}&\multicolumn{1}{c}{3.27\%}&\multicolumn{1}{c}{3.04\%}&\multicolumn{1}{c}{1.51\%}&\multicolumn{1}{c}{2.32\%}\\
\noalign{\smallskip}\hline\noalign{\smallskip}
\multicolumn{1}{c}{2}&\multicolumn{1}{c}{131.66}&\multicolumn{1}{c}{78.95}&\multicolumn{1}{c}{113.25}&\multicolumn{1}{c}{82.07}&\multicolumn{1}{c}{127.44}&\multicolumn{1}{c}{72.79}\\
\multicolumn{1}{c}{(III)}&\multicolumn{1}{c}{$\pm$1.28}&\multicolumn{1}{c}{$\pm$1.01}&\multicolumn{1}{c}{$\pm$4.28}&\multicolumn{1}{c}{$\pm$2.49}&\multicolumn{1}{c}{$\pm$2.78}&\multicolumn{1}{c}{$\pm$1.49}\\
\multicolumn{1}{c}{}&\multicolumn{1}{c}{0.97\%}&\multicolumn{1}{c}{1.28\%}&\multicolumn{1}{c}{3.78\%}&\multicolumn{1}{c}{3.03\%}&\multicolumn{1}{c}{2.18\%}&\multicolumn{1}{c}{2.05\%}\\
\noalign{\smallskip}\hline\noalign{\smallskip}
\multicolumn{1}{c}{3}&\multicolumn{1}{c}{139.28}&\multicolumn{1}{c}{75.03}&\multicolumn{1}{c}{157.41}&\multicolumn{1}{c}{74.18}&\multicolumn{1}{c}{153.29}&\multicolumn{1}{c}{73.37}\\
\multicolumn{1}{c}{(II)}&\multicolumn{1}{c}{$\pm$3.02}&\multicolumn{1}{c}{$\pm$1.79}&\multicolumn{1}{c}{$\pm$2.34}&\multicolumn{1}{c}{1.54}&\multicolumn{1}{c}{$\pm$3.46}&\multicolumn{1}{c}{$\pm$2.04}\\
\multicolumn{1}{c}{}&\multicolumn{1}{c}{2.17\%}&\multicolumn{1}{c}{2.39\%}&\multicolumn{1}{c}{1.48\%}&\multicolumn{1}{c}{2.07\%}&\multicolumn{1}{c}{2.26\%}&\multicolumn{1}{c}{2.79\%}\\
\noalign{\smallskip}\hline\noalign{\smallskip}
\multicolumn{1}{c}{4}&\multicolumn{1}{c}{122.36}&\multicolumn{1}{c}{63.36}&\multicolumn{1}{c}{122.64}&\multicolumn{1}{c}{55.49}&\multicolumn{1}{c}{127.74}&\multicolumn{1}{c}{58.30}\\
\multicolumn{1}{c}{(III)}&\multicolumn{1}{c}{$\pm$1.89}&\multicolumn{1}{c}{$\pm$0.56}&\multicolumn{1}{c}{$\pm$2.41}&\multicolumn{1}{c}{1.09}&\multicolumn{1}{c}{$\pm$1.73}&\multicolumn{1}{c}{$\pm$0.84}\\
\multicolumn{1}{c}{}&\multicolumn{1}{c}{1.55\%}&\multicolumn{1}{c}{0.89\%}&\multicolumn{1}{c}{1.96\%}&\multicolumn{1}{c}{1.97\%}&\multicolumn{1}{c}{1.35\%}&\multicolumn{1}{c}{1.45\%}\\
\noalign{\smallskip}\hline\noalign{\smallskip}
\multicolumn{1}{c}{5}&\multicolumn{1}{c}{158.01}&\multicolumn{1}{c}{84.11}&\multicolumn{1}{c}{179.73}&\multicolumn{1}{c}{80.65}&\multicolumn{1}{c}{175.68}&\multicolumn{1}{c}{73.17}\\
\multicolumn{1}{c}{(II)}&\multicolumn{1}{c}{$\pm$5.19}&\multicolumn{1}{c}{$\pm$0.74}&\multicolumn{1}{c}{$\pm$4.32}&\multicolumn{1}{c}{3.46}&\multicolumn{1}{c}{$\pm$4.64}&\multicolumn{1}{c}{$\pm$2.78}\\
\multicolumn{1}{c}{}&\multicolumn{1}{c}{3.28\%}&\multicolumn{1}{c}{0.88\%}&\multicolumn{1}{c}{2.40\%}&\multicolumn{1}{c}{4.29\%}&\multicolumn{1}{c}{2.64\%}&\multicolumn{1}{c}{3.80\%}\\
\noalign{\smallskip}\hline\noalign{\smallskip}
\multicolumn{1}{c}{6}&\multicolumn{1}{c}{143.16}&\multicolumn{1}{c}{76.48}&\multicolumn{1}{c}{152.47}&\multicolumn{1}{c}{71.45}&\multicolumn{1}{c}{156.64}&\multicolumn{1}{c}{71.28}\\
\multicolumn{1}{c}{(II)}&\multicolumn{1}{c}{$\pm$2.34}&\multicolumn{1}{c}{$\pm$2.86}&\multicolumn{1}{c}{$\pm$2.95}&\multicolumn{1}{c}{1.91}&\multicolumn{1}{c}{$\pm$2.97}&\multicolumn{1}{c}{$\pm$2.34}\\
\multicolumn{1}{c}{}&\multicolumn{1}{c}{1.64\%}&\multicolumn{1}{c}{3.74\%}&\multicolumn{1}{c}{1.94\%}&\multicolumn{1}{c}{2.68\%}&\multicolumn{1}{c}{1.90\%}&\multicolumn{1}{c}{3.28\%}\\
\noalign{\smallskip}\hline\noalign{\smallskip}
\multicolumn{1}{c}{7}&\multicolumn{1}{c}{149.05}&\multicolumn{1}{c}{69.16}&\multicolumn{1}{c}{155.87}&\multicolumn{1}{c}{68.46}&\multicolumn{1}{c}{156.16}&\multicolumn{1}{c}{65.15}\\
\multicolumn{1}{c}{(II)}&\multicolumn{1}{c}{$\pm$5.29}&\multicolumn{1}{c}{$\pm$2.28}&\multicolumn{1}{c}{$\pm$4.78}&\multicolumn{1}{c}{1.67}&\multicolumn{1}{c}{$\pm$4.65}&\multicolumn{1}{c}{$\pm$2.63}\\
\multicolumn{1}{c}{}&\multicolumn{1}{c}{3.55\%}&\multicolumn{1}{c}{3.30\%}&\multicolumn{1}{c}{3.07\%}&\multicolumn{1}{c}{2.45\%}&\multicolumn{1}{c}{2.98\%}&\multicolumn{1}{c}{4.04\%}\\
\noalign{\smallskip}\hline\noalign{\smallskip}
\multicolumn{1}{c}{8}&\multicolumn{1}{c}{119.96}&\multicolumn{1}{c}{67.01}&\multicolumn{1}{c}{129.24}&\multicolumn{1}{c}{65.94}&\multicolumn{1}{c}{136.29}&\multicolumn{1}{c}{65.13}\\
\multicolumn{1}{c}{(I)}&\multicolumn{1}{c}{$\pm$3.51}&\multicolumn{1}{c}{$\pm$1.58}&\multicolumn{1}{c}{$\pm$2.75}&\multicolumn{1}{c}{1.95}&\multicolumn{1}{c}{$\pm$2.12}&\multicolumn{1}{c}{$\pm$1.11}\\
\multicolumn{1}{c}{}&\multicolumn{1}{c}{2.92\%}&\multicolumn{1}{c}{2.36\%}&\multicolumn{1}{c}{2.13\%}&\multicolumn{1}{c}{2.96\%}&\multicolumn{1}{c}{1.55\%}&\multicolumn{1}{c}{1.70\%}\\
\noalign{\smallskip}\hline\noalign{\smallskip}
\multicolumn{1}{c}{9}&\multicolumn{1}{c}{161.22}&\multicolumn{1}{c}{74.52}&\multicolumn{1}{c}{163.07}&\multicolumn{1}{c}{69.26}&\multicolumn{1}{c}{154.79}&\multicolumn{1}{c}{68.09}\\
\multicolumn{1}{c}{(IV)}&\multicolumn{1}{c}{$\pm$3.61}&\multicolumn{1}{c}{$\pm$2.12}&\multicolumn{1}{c}{$\pm$3.35}&\multicolumn{1}{c}{1.48}&\multicolumn{1}{c}{$\pm$10.63}&\multicolumn{1}{c}{$\pm$6.02}\\
\multicolumn{1}{c}{}&\multicolumn{1}{c}{2.24\%}&\multicolumn{1}{c}{2.85\%}&\multicolumn{1}{c}{2.05\%}&\multicolumn{1}{c}{2.14\%}&\multicolumn{1}{c}{6.87\%}&\multicolumn{1}{c}{8.84\%}\\
\noalign{\smallskip}\hline\noalign{\smallskip}
\multicolumn{1}{c}{10}&\multicolumn{1}{c}{135.91}&\multicolumn{1}{c}{73.69}&\multicolumn{1}{c}{140.07}&\multicolumn{1}{c}{70.72}&\multicolumn{1}{c}{157.84}&\multicolumn{1}{c}{72.85}\\
\multicolumn{1}{c}{(III)}&\multicolumn{1}{c}{$\pm$2.72}&\multicolumn{1}{c}{$\pm$0.95}&\multicolumn{1}{c}{$\pm$4.81}&\multicolumn{1}{c}{2.07}&\multicolumn{1}{c}{$\pm$2.22}&\multicolumn{1}{c}{$\pm$0.95}\\
\multicolumn{1}{c}{}&\multicolumn{1}{c}{2\%}&\multicolumn{1}{c}{1.29\%}&\multicolumn{1}{c}{3.43\%}&\multicolumn{1}{c}{2.93\%}&\multicolumn{1}{c}{1.41\%}&\multicolumn{1}{c}{1.30\%}\\
\noalign{\smallskip}\hline\noalign{\smallskip}
\multicolumn{1}{c}{11}&\multicolumn{1}{c}{177.26}&\multicolumn{1}{c}{74.32}&\multicolumn{1}{c}{177.21}&\multicolumn{1}{c}{72.86}&\multicolumn{1}{c}{183.60}&\multicolumn{1}{c}{73.26}\\
\multicolumn{1}{c}{(III)}&\multicolumn{1}{c}{$\pm$5.75}&\multicolumn{1}{c}{$\pm$2.82}&\multicolumn{1}{c}{$\pm$2.83}&\multicolumn{1}{c}{1.17}&\multicolumn{1}{c}{$\pm$3.28}&\multicolumn{1}{c}{$\pm$1.35}\\
\multicolumn{1}{c}{}&\multicolumn{1}{c}{3.24\%}&\multicolumn{1}{c}{3.80\%}&\multicolumn{1}{c}{1.60\%}&\multicolumn{1}{c}{1.61\%}&\multicolumn{1}{c}{1.79\%}&\multicolumn{1}{c}{1.85\%}\\
\noalign{\smallskip}\hline\noalign{\smallskip}
\multicolumn{1}{c}{12}&\multicolumn{1}{c}{137.46}&\multicolumn{1}{c}{71.90}&\multicolumn{1}{c}{152.44}&\multicolumn{1}{c}{69}&\multicolumn{1}{c}{145.57}&\multicolumn{1}{c}{65.81}\\
\multicolumn{1}{c}{(II)}&\multicolumn{1}{c}{$\pm$4.11}&\multicolumn{1}{c}{$\pm$2.03}&\multicolumn{1}{c}{$\pm$3.43}&\multicolumn{1}{c}{2.69}&\multicolumn{1}{c}{$\pm$3.01}&\multicolumn{1}{c}{$\pm$1.95}\\
\multicolumn{1}{c}{}&\multicolumn{1}{c}{3\%}&\multicolumn{1}{c}{2.82\%}&\multicolumn{1}{c}{2.25\%}&\multicolumn{1}{c}{3.90\%}&\multicolumn{1}{c}{2.07\%}&\multicolumn{1}{c}{2.97\%}\\
\noalign{\smallskip}\hline
\end{tabular}
\end{scriptsize}
\end{center}
\label{Tab4}
\end{table}

\begin{table}[ht!]
\caption{Errors ([mmHg]) in simulated central systolic (sys), diastolic (dia), mean (mean) and pulse (pp) pressures through both the general (superscript G) and patient-specific (superscript PS) models, evaluated with respect to the mean values of the same invasive pressures along the ascending aorta (AA). Moduli of mean errors $\pm$ standard deviation are reported in bold in the last row.}
\begin{center}
\begin{scriptsize}
\begin{tabular}{c@{\quad}r@{\quad}r@{\quad}r@{\quad}r@{\quad}r@{\quad}r@{\quad}r@{\quad}r}
\hline\noalign{\smallskip}
\multirow{3}{*}{Patient}&\multicolumn{4}{c}{General model}&\multicolumn{4}{c}{Patient-specific model}\\
\cline{2-9}
&\multicolumn{1}{c}{sys$^{G}$}&\multicolumn{1}{c}{dia$^{G}$}&\multicolumn{1}{c}{mean$^{G}$}&\multicolumn{1}{c}{pp$^{G}$}&\multicolumn{1}{c}{sys$^{PS}$}&\multicolumn{1}{c}{dia$^{PS}$}&\multicolumn{1}{c}{mean$^{PS}$}&\multicolumn{1}{c}{pp$^{PS}$}\\
\noalign{\smallskip}\hline\noalign{\smallskip}
\multicolumn{1}{c}{1}&\multicolumn{1}{c}{0.19}&\multicolumn{1}{c}{16.08}&\multicolumn{1}{c}{10.78}&\multicolumn{1}{c}{-15.89}&\multicolumn{1}{c}{0.14}&\multicolumn{1}{c}{-3.23}&\multicolumn{1}{c}{-2.10}&\multicolumn{1}{c}{3.37}\\
\multicolumn{1}{c}{2}&\multicolumn{1}{c}{-16.93}&\multicolumn{1}{c}{-4.66}&\multicolumn{1}{c}{-8.75}&\multicolumn{1}{c}{-15.89}&\multicolumn{1}{c}{-2.25}&\multicolumn{1}{c}{-2.71}&\multicolumn{1}{c}{-2.56}&\multicolumn{1}{c}{0.45}\\
\multicolumn{1}{c}{3}&\multicolumn{1}{c}{-24.55}&\multicolumn{1}{c}{-0.74}&\multicolumn{1}{c}{-8.68}&\multicolumn{1}{c}{-23.81}&\multicolumn{1}{c}{2.90}&\multicolumn{1}{c}{5.14}&\multicolumn{1}{c}{4.39}&\multicolumn{1}{c}{-2.23}\\
\multicolumn{1}{c}{4}&\multicolumn{1}{c}{-7.63}&\multicolumn{1}{c}{10.90}&\multicolumn{1}{c}{4.74}&\multicolumn{1}{c}{-18.56}&\multicolumn{1}{c}{-7.53}&\multicolumn{1}{c}{-10.16}&\multicolumn{1}{c}{-9.28}&\multicolumn{1}{c}{2.63}\\
\multicolumn{1}{c}{5}&\multicolumn{1}{c}{-43.28}&\multicolumn{1}{c}{-9.82}&\multicolumn{1}{c}{-20.97}&\multicolumn{1}{c}{-33.46}&\multicolumn{1}{c}{1.84}&\multicolumn{1}{c}{9.80}&\multicolumn{1}{c}{7.15}&\multicolumn{1}{c}{-7.96}\\
\multicolumn{1}{c}{6}&\multicolumn{1}{c}{-28.43}&\multicolumn{1}{c}{-2.19}&\multicolumn{1}{c}{-10.94}&\multicolumn{1}{c}{-26.24}&\multicolumn{1}{c}{-4.94}&\multicolumn{1}{c}{-6.55}&\multicolumn{1}{c}{-6.01}&\multicolumn{1}{c}{1.60}\\
\multicolumn{1}{c}{7}&\multicolumn{1}{c}{-34.32}&\multicolumn{1}{c}{5.13}&\multicolumn{1}{c}{-8.02}&\multicolumn{1}{c}{-39.45}&\multicolumn{1}{c}{2.62}&\multicolumn{1}{c}{-1.38}&\multicolumn{1}{c}{-0.045}&\multicolumn{1}{c}{3.99}\\
\multicolumn{1}{c}{8}&\multicolumn{1}{c}{-5.23}&\multicolumn{1}{c}{7.28}&\multicolumn{1}{c}{3.11}&\multicolumn{1}{c}{-12.51}&\multicolumn{1}{c}{-4.80}&\multicolumn{1}{c}{-1.62}&\multicolumn{1}{c}{-2.68}&\multicolumn{1}{c}{-3.18}\\
\multicolumn{1}{c}{9}&\multicolumn{1}{c}{-46.49}&\multicolumn{1}{c}{-0.23}&\multicolumn{1}{c}{-15.65}&\multicolumn{1}{c}{-46.26}&\multicolumn{1}{c}{-8.55}&\multicolumn{1}{c}{-14.56}&\multicolumn{1}{c}{-12.56}&\multicolumn{1}{c}{6.01}\\
\multicolumn{1}{c}{10}&\multicolumn{1}{c}{-21.18}&\multicolumn{1}{c}{0.60}&\multicolumn{1}{c}{-6.66}&\multicolumn{1}{c}{-21.78}&\multicolumn{1}{c}{-1.60}&\multicolumn{1}{c}{-3.07}&\multicolumn{1}{c}{-2.58}&\multicolumn{1}{c}{1.48}\\
\multicolumn{1}{c}{11}&\multicolumn{1}{c}{-62.53}&\multicolumn{1}{c}{-0.03}&\multicolumn{1}{c}{-20.86}&\multicolumn{1}{c}{-62.50}&\multicolumn{1}{c}{-8.34}&\multicolumn{1}{c}{-10.36}&\multicolumn{1}{c}{-9.69}&\multicolumn{1}{c}{2.02}\\
\multicolumn{1}{c}{12}&\multicolumn{1}{c}{-22.73}&\multicolumn{1}{c}{2.39}&\multicolumn{1}{c}{-5.98}&\multicolumn{1}{c}{-25.12}&\multicolumn{1}{c}{-5.56}&\multicolumn{1}{c}{1.69}&\multicolumn{1}{c}{-0.73}&\multicolumn{1}{c}{-7.25}\\
&\multicolumn{1}{c}{\textbf{26.12}}&\multicolumn{1}{c}{\textbf{5}}&\multicolumn{1}{c}{\textbf{10.43}}&\multicolumn{1}{c}{\textbf{28.46}}&\multicolumn{1}{c}{\textbf{4.26}}&\multicolumn{1}{c}{\textbf{5.86}}&\multicolumn{1}{c}{\textbf{4.98}}&\multicolumn{1}{c}{\textbf{3.51}}\\
&\multicolumn{1}{c}{\textbf{$\pm$18.29}}&\multicolumn{1}{c}{\textbf{$\pm$5.10}}&\multicolumn{1}{c}{\textbf{$\pm$5.87}}&\multicolumn{1}{c}{\textbf{$\pm$14.68}}&\multicolumn{1}{c}{\textbf{$\pm$2.81}}&\multicolumn{1}{c}{\textbf{$\pm$4.38}}&\multicolumn{1}{c}{\textbf{$\pm$3.95}}&\multicolumn{1}{c}{\textbf{$\pm$2.38}}\\
\noalign{\smallskip}\hline
\end{tabular}
\end{scriptsize}
\end{center}
\label{Tab5}
\end{table}

\end{document}